\documentclass[aps,
		prd,
		reprint,
        superscriptaddress,
        shortbibliography,
		  nofootinbib,
		floatfix,
		notitlepage, onecolumn
		]{revtex4-1}
\pdfoutput=1

\usepackage{amsmath}
\usepackage{mathtools}
\usepackage{amssymb}
\usepackage{slashed}
\usepackage{color}
\usepackage{xcolor}
\usepackage{hyperref}
\usepackage{braket}
\hypersetup{colorlinks=true,linkcolor=blue} 
\usepackage{comment}

\newcommand{\be}{\begin{equation}}
\newcommand{\ee}{\end{equation}}
\newcommand{\bi}{\begin{enumerate}}
\newcommand{\ei}{\end{enumerate}}
\newcommand{\ud}{{\mathrm{d}}}
\hfuzz=\maxdimen \tolerance=10000
\vfuzz=\maxdimen \tolerance=10000
\newcommand{\LCm}{{\scriptscriptstyle -}}
\newcommand{\LCp}{{\scriptscriptstyle +}}

\newcommand{\LCperp}{{\scriptscriptstyle \perp}}
\renewcommand{\j}{\theta}

\newcommand{\llangle}{\langle\!\langle}
\newcommand{\rrangle}{\rangle\!\rangle}
\newcommand{\lDot}{{}^{\bullet}\!}

\newcommand{\rDot}{\!{}^{\bullet}}

\def\ddel{{}^\bullet\! \Delta}
\def\deld{\Delta^{\hskip -.5mm \bullet}}
\def\dddel{{}^{\bullet \bullet} \! \Delta}
\def\ddeld{{}^{\bullet}\! \Delta^{\hskip -.5mm \bullet}}

\newcommand{\Pn}{\bar{\mathfrak{P}}}

\newcommand{\e}{\mathrm{e}}

\begin{document}

\title{Master Formulae for $N$-photon tree level amplitudes in plane wave backgrounds}
\author{Patrick Copinger}
\email{patrick.copinger@plymouth.ac.uk}
\affiliation{Centre for Mathematical Sciences, University of Plymouth, Plymouth, PL4 8AA, UK}
\author{James P. Edwards}
\email{james.p.edwards@plymouth.ac.uk}
\affiliation{Centre for Mathematical Sciences, University of Plymouth, Plymouth, PL4 8AA, UK}
\author{Anton Ilderton}
\email{anton.ilderton@ed.ac.uk}
\affiliation{Higgs Centre, School of Physics and Astronomy, University of Edinburgh, EH9 3FD, UK}
\author{Karthik Rajeev}
\email{karthik.rajeev@ed.ac.uk}
\affiliation{Higgs Centre, School of Physics and Astronomy, University of Edinburgh, EH9 3FD, UK}

\begin{abstract}
The presence of strong electromagnetic fields adds huge complexity to QED Feynman diagrams, such that new methods are required to calculate higher-loop and higher-multiplicity scattering amplitudes. Here we use the worldline formalism to present `Master Formulae' for all tree level amplitudes of two massive particles and an arbitrary number of photons, in a plane wave background, in both scalar and spinor QED. The plane wave is treated without approximation throughout, meaning in particular that our formulae are valid in the strong-field regime of current theoretical and experimental interest. We check our results against literature expressions obtainable at low multiplicity via direct Feynman diagram calculations.
\end{abstract}

\maketitle
\onecolumngrid

\section{Introduction}
%
Strong fields can generate nonlinear and non-perturbative effects in particle interactions. Strong electromagnetic fields may be generated terrestrially by several means, including by ultra-intense lasers~\cite{Strickland:1985gxr,Nobel}. QED processes in the presence of these fields acquire an intensity-dependence characterised by a coupling which typically exceeds unity, and which must therefore be treated without recourse to perturbation theory. Several upcoming experiments aim to observe nonlinear effects in the scattering of electrons~\cite{ELI,Abramowicz:2021zja,Clarke:2022rbd} and photons~\cite{Schlen,Karbstein:2021otv} on intense lasers.

The standard theory approach to `strong field QED' is based on the Furry expansion, or background field perturbation theory. The strong (e.g.~laser) field is described as a fixed background, the coupling of which to matter is treated exactly. Interactions between particles scattering on this background are then treated in perturbation theory as usual, see~\cite{Fedotov:2022ely} for a recent review.  There are however several topics in strong field QED which require the development of new theoretical methods.

First, the majority of progress to date has been made for the special, highly symmetric laser model of a plane wave background, for which the Furry expansion can be practically realised. It is a long-standing challenge to account analytically for realistic pulse geometry, and the new phenomenology this brings~\cite{Fedotov:2022ely}. Second, while plane wave results can be extended to realistic fields via local approximations (e.g.~\cite{DiPiazza:2018bfu,Ilderton:2018nws,Heinzl:2020ynb}), and so implemented in numerical codes, those codes must still be benchmarked against theory. This has been performed for first-order (i.e.~low multiplicity) processes, but benchmarking higher-order processes is made challenging by, in part, a lack of analytic results; the state-of-the-art in the plane wave model is, at tree level, only \emph{four}-point scattering. Third, if we consider higher loop corrections, it has been conjectured~\cite{RitusRN1,Narozhnyi:1980dc,Fedotov:2016afw} that at very high background field strengths the loop expansion must be resummed in order to provide reliable physical predictions (at least in the low frequency, `constant crossed field' limit). Doing so is a formidable challenge~\cite{Heinzl:2021mji,Mironov:2022jbg,Torgrimsson:2021wcj}.

To attack these problems one can use approximations that do not rely on weak coupling~\cite{DiPiazza:2016maj}, develop exactly solvable models which capture some physics of interest~\cite{Heinzl:2017zsr}, or use alternative methods to simplify Furry-picture quantities. One potential method is the worldline formalism, which casts QFT in terms of path integrals over relativistic point particle trajectories. %
    Its roots can be traced back to Feynman~\cite{Feyn1,Feyn2}, though its use as a serious alternative to the standard QFT formalism was first advocated by Strassler~\cite{Strass1}, following~\cite{Bern:1990cu,Bern:1991aq}. One of the main advantages of the worldline approach is that it automatically sums over all Feynman diagrams which contribute at fixed multiplicity and loop order, thus greatly simplifying the combinatorics which comes with higher numbers of scatterers and/or loops.

The worldline formalism was initially developed for one-loop (and then higher loop) processes in vacuum and in background fields, and a common output of the approach is `Master Formulae'; these are \emph{all}-multiplicity formulae for correlation functions of a chosen set of fields, at fixed loop order. Such Master Formulae, which would be extremely challenging to reproduce using Feynman diagrams, have been obtained for processes in vacuum~\cite{Strass1,Schmidt:1993rk,103}, in constant electromagnetic backgrounds~\cite{Shaisultanov:1995tm,Adler:1996cja,Reuter:1996zm,Dittrich:2000wz,Schubert:2000yt,McKeon:1994hd} and in plane wave backgrounds~\cite{Edwards:2021vhg, Edwards:2021uif,Schubert:2023gsl}. 
The worldline approach has also been applied to the calculation of effective actions in background fields via numerical implementations~\cite{Gies:2001tj}, the Casimir effect~\cite{Gies:2003cv}, vacuum birefringence~\cite{AntonPlane}, tadpole corrections~\cite{Ahmadiniaz:2019nhk, Ahmadiniaz:2017rrk, Edwards:2017bte}, and nonlinear Breit-Wheeler pair production~\cite{DegliEsposti:2021its}.
A long-standing focus of the approach has been the investigation of non-perturbative effects via worldline instantons~\cite{Affleck:1981bma,Srinivasan:1998ty,Kim:2000un,Dunne:2005sx,Dunne:2006st,Dumlu:2011cc,Ilderton:2015qda}. For reviews see~\cite{ChrisRev, UsRep}.

Only recently has much attention been paid to worldline Master Formulae for processes with external matter lines, or processes at tree level \cite{Ahmadiniaz:2020wlm, Ahmadiniaz:2021gsd, Bhattacharya:2017wlw, Ahmad:2016vvw, Ahmadiniaz:2015xoa, Corradini:2020prz}. Furthermore, while external photon lines typically appear in the worldline formalism already LSZ-amputated, matter lines do not, and it has not yet been fully established how one should perform the required LSZ amputation which turns correlation functions into amplitudes.

We fill in some missing pieces of this puzzle in this paper, which is organised as follows.
In Sect.~\ref{sect:correlators} we construct worldline Master Formulae for all tree level $N+2$-point correlation functions describing the emission of $N$ photons from a massive particle in a background plane wave, in both scalar and spinor QED. In Sect.~\ref{sect:LSZ} we turn to the LSZ amputation of the master formula, converting it into an all-multiplicity formula for the corresponding $N$-photon emission/absorption amplitudes from a massive particle in a plane wave background.
Example calculations in which we compare with known literature results at low multiplicity are  presented in Sect.~\ref{sect:examples}. We conclude in Sect.~\ref{sect:conclusions}. The appendices contain additional checks on our results.

\paragraph*{Conventions:} We set $\hbar=c=1$. We work throughout in Minkowski space with lightfront coordiantes, so that $\ud s^2 = \ud x^\LCp \ud x^\LCm - \ud x^\LCperp \ud x^\LCperp$ where $x^\LCperp = (x^1,x^2)$ are the `transverse' directions. We introduce a null vector $n_\mu$ which projects onto the `lightfront time' direction, that is $n\cdot x = x^\LCp$. The covariant derivative is $D_\mu = \partial_\mu + i e A_\mu$.

\section{Master Formulae for $2+N$--point tree level correlators in plane wave backgrounds}
\label{sect:correlators}

\begin{figure}[t!]
\includegraphics[width=0.7\textwidth]{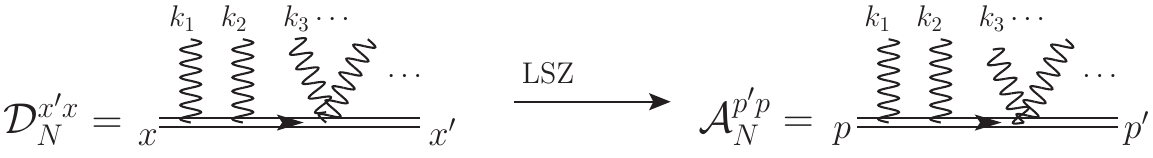}
\caption{\label{Fig1} We consider tree level scattering amplitudes of 2 massive charges and $N$ photons, as illustrated on the \emph{right} (for scalar QED). The double line represents the presence of a plane wave background, the coupling to which is treated exactly. Amplitudes are obtained by LSZ reduction of the corresponding correlation functions. In the worldline approach, a natural starting pointing is the partially amputated correlator, or `dressed propagator', in which the photons are already reduced out, but the matter fields are not. This is illustrated on the \emph{left}. Thus LSZ reduction is still required for the external matter lines.}
\end{figure}

The goal of this section is to write down and evaluate the worldline path integral Master Formulae for tree level \emph{correlation functions} of $N$ photons and two charged particles in the presence of a plane wave background, valid for arbitrary $N$. We will do this in both scalar and spinor QED. 

Our plane wave background may be described by the potential $e A_\mu(x) = a_\mu(x^\LCp) = \delta_\mu^\LCperp a_\LCperp (x^\LCp)$, a transverse function of lightfront time  $x^{\LCp}$. We may always choose $a_\LCperp(-\infty)=0$, but then $a_\LCperp(\infty) \eqqcolon a_\LCperp^\infty$ is in general non-zero (and carries an electromagnetic memory effect~\cite{Dinu:2012tj,Bieri:2013hqa,Cristofoli:2022phh}.). The corresponding field strength is $f_{\mu\nu}(x^\LCp) = n_\mu a'_\nu(x^\LCp) - n_\nu a'_\mu(x^\LCp)$, where a prime denotes an $x^\LCp$-derivative.

\subsection{Scalar QED}\label{subsec:SQED}
%
In the Master Formulae we derive in this section, the $N$ external photons will be LSZ-amputated, but the matter lines not, and thus our correlation functions carry spacetime indices $x$ and $x'$, as well as a dependence on the $N$ photon momenta $\{k_i\}$ and polarisations $\{\varepsilon_i\}$. We hide the latter dependencies, denoting the partially reduced correlators, or `dressed propagators' as they are called in the worldline literature, by $\mathcal{D}_N^{x'x}$; see Fig.~\ref{Fig1}. We take all photons to be \emph{outgoing}; other configurations are trivially obtained by sending $k\to-k$.

The worldline representation of such correlation functions is given in terms of a path integral over relativistic point particle trajectories, denoted $x^{\mu}(\tau)$ with $\tau$ the proper time of the trajectory.
The trajectories obey Dirichlet boundary conditions $x^\mu(T) = x^{\prime \mu}$, $x^\mu(0)=x^\mu$, corresponding to the spacetime dependence of the dressed propagator. The trajectories have length $T$, which is ultimately also integrated out, respecting reparameterisation invariance of the path integral~\cite{Polyakov:1987ez,Mansfield:1990tu}. To write down this path integral, we start from the worldline action that minimally couples a relativistic point particle to an arbitrary gauge field $A_{\mu}$, namely
\begin{equation}
     S_{\mathrm{WL}}[x(\tau), A] = -\int_{0}^{T}\!\ud\tau\,\Big[\frac{\dot{x}^{2}}{4}+e{A(x(\tau))\cdot \dot{x}(\tau)} \Big]\;,
\end{equation}
where over-dots denote proper time derivatives, and where the unusual normalisation of the kinetic term has become standard in the worldline literature, so we preserve it here. $S_{\mathrm{WL}}$ enters the path integral for the scalar field propagator, call it $\mathcal{D}^{x'x}$, via
\be\label{propagator_pos_pi}
    \mathcal{D}^{x'x} = 
    \int_{0}^{\infty}\ud T \,\e^{-im^{2} T}\,
\int^{x(T)=x'}_{x(0)=x}\mathcal{D}x(\tau)\, \e^{iS_{\mathrm{WL}}[x(\tau), A]}\;.
\ee
Note that $A_\mu$ is not integrated over, rather it appears as a given field -- it is well known (see, for example \cite{Itzykson:1980rh}) that correlation functions with $N$ external photons \emph{in vacuum} can be extracted from (\ref{propagator_pos_pi}) by fixing $A_\mu$ to be a sum over asymptotic photon wavefunctions with momenta $k_{i}$ and polarisations $\varepsilon_{i}$:
\begin{equation}
    \label{Total-A2}
    A_\mu(x) \to A^{\gamma}_{\mu}(x) = \sum_{i=1}^{N}\varepsilon_{\mu\, i}\e^{i k_{i}\cdot x}\,,
\end{equation}
and then expanding the dressed propagator (\ref{propagator_pos_pi}) to multi-linear order in the polarisation vectors. The additional complication here is the presence of the background gauge potential in (\ref{eqSB}). This is however easily included; we simply split the gauge field into a semi-classical part representing the plane wave background and a `quantised' part representing scattering photons: 
\begin{equation}
    \label{Total-A1}
    eA_{\mu}(x) \to a_{\mu}(x) + eA^{\gamma}_{\mu}(x)\,.
\end{equation}
Inserting this into (\ref{propagator_pos_pi}) and expanding to multi-linear order, the path integral to be performed is
\be \label{propagator_pos_def}
    \mathcal{D}^{x'x}_N = 
    (-ie)^{N}
    \int_{0}^{\infty}\ud T \,\e^{-im^{2} T}\,
\int^{x(T)=x'}_{x(0)=x}\hspace{-1em}\mathcal{D}x(\tau)\, \e^{iS_{\mathrm{B}}[x(\tau), a]}\prod_{i=1}^{N} V^{x'x}[\varepsilon_{i}, k_i]\;,
\ee
in which the weight is now given by the reduced action
\be
    S_{\mathrm{B}}[x(\tau), a] = -\int_{0}^{T}\!\ud\tau\, \Big[\frac{\dot{x}^{2}}{4}+{a(x(\tau))\cdot \dot{x}(\tau)} \Big]\;,
    \label{eqSB}
\ee
while the $N$ external photons appear (following the expansion to multi-linear order) through the vertex functions
\begin{equation}
    V^{x'x}[\varepsilon,k]\coloneqq \int_{0}^{T}d\tau\,\varepsilon\cdot\dot{x}(\tau)\, \e^{ik\cdot x(\tau)}\,.
\end{equation}
(We leave implicit a causal and IR convergence factor $\exp(-\epsilon T)$ under the $\ud T$ integral in (\ref{propagator_pos_def}).)

Our task is to evaluate the integrals in (\ref{propagator_pos_def}).
Let us first consider the $x^\mu$ integrals, and in particular the Dirichlet BCs.
To deal with these we follow the standard procedure used for the evaluation of such integrals in vacuum, and expand $x^\mu(\tau)$ into a straight line trajectory and a fluctuation $q(\tau)$ according to
\begin{equation}
    x^{\mu}(\tau)= x^{\mu}+z^\mu\frac{\tau}{T}+q^{\mu}(\tau)\,, \qquad z^\mu := x^{\prime\mu}-x^{\mu} \;.
\end{equation}
The fluctuation must satisfy the homogeneous Dirichlet BCs $q(0) = q(T) = 0$ (with measure $\mathcal{D}x(\tau) \rightarrow \mathcal{D}q(\tau)$). For the analogue problem in vacuum ($a(x^{\LCp}) \rightarrow 0$) the path integral is Gaussian in $q_\mu$ and can thus be computed analytically\footnote{This is also the case for a constant background in Fock-Schwinger gauge~\cite{Ahmad:2016vvw}.}. Here, however, the fluctuation appears \emph{inside} the background field $a(x^\LCp(\tau)) = a(x^\LCp+z^\LCp \tau/T+q^\LCp)$, and this has an arbitrary functional form.  At first glance this seems to destroy the Gaussianity of the path integral, and prohibit its evaluation. However, it has been shown for one-loop photon-scattering processes (meaning no external matter lines, and a path integral with periodic rather than Dirichlet BCs) that the properties of the plane wave background mean the integral is still effectively Gaussian~\cite{AntonPlane,Edwards:2021vhg,*Edwards:2021uif}. It is thus crucial to demonstrate that the hidden Gaussianity of the path integral is also present here.

To do so we follow the approach of~\cite{Schubert:2023gsl}, introducing a Lagrange multiplier $\chi(\tau)$ and auxiliary field $\xi(\tau)$  into the path integral through the equality
\begin{equation}
\e^{-i\int \ud\tau\, a(x^\LCp(\tau)) \cdot \dot{q}}
=
\e^{-i\int \ud\tau\, a(x^\LCp+z^\LCp\frac{\tau}{T}+q^\LCp) \cdot \dot{q}}
=
\int\!\mathcal{D}\xi\mathcal{D}\chi\,
\e^{i\int \ud\tau\, \big[\chi(\xi-q^\LCp)-a(x^\LCp+z^\LCp\frac{\tau}{T}+\xi)\cdot\dot{q}\big]}\,.
\end{equation}
These auxiliary integrals render that over $q(\tau)$ to be Gaussian. The crucial point, as we show below, is that after evaluating the $q$-integral, the remaining integrals over $\xi$ and $\chi$ can still be evaluated, for a plane wave background.

We now compute the fluctuation integral. As is usual in this `string-inspired' approach, it is convenient to manipulate the vertex operators  as follows. We exponentiate the polarisation-dependent factor, so that it appears linearly in an exponent in the operator, with the understanding that the result should later be expanded to linear order in (each of) the $\varepsilon_{i}$, so we write
\be
    V^{x'x}[\varepsilon,k] \to \int_{0}^{T}\!\ud\tau\, \e^{ik\cdot x + \varepsilon\cdot\dot{x}}\Big|_{\mathrm{lin. }\,\varepsilon} \;.
\ee
The result of this is that all dependence on the particle trajectory $x(\tau)$, or rather the fluctuation $q(\tau)$ to be integrated out, now appears \emph{linearly} under the path integral. The integrals to be evaluated are now
\begin{equation*}
   \mathcal{D}_N^{x'x} =
   (-ie)^{N}\!
  {
  \int_{0}^{\infty}\!\ud T \,\e^{-im^{2} T-i \frac{z^{2}}{4T}}
  }
   \prod_{i = 1}^{N} \int_{0}^{T}\!\ud \tau_{i}\,
   \e^{\,\sum\limits_{j =1}^{N} ik_{j} \cdot (x + z\frac{\tau_{j}}{T})  + \varepsilon_{j} \cdot \frac{z}{T}} \int\! \mathcal{D}\xi\, \mathcal{D}\chi \, \int_{q(0) = 0}^{q(T) = 0} \!\mathcal{D}q(\tau)\,
 \e^{i\int \ud\tau[-\frac{\dot{q}^{2}}{4}-\mathcal{J}\cdot q]} \Big|_{\mathrm{lin. }\; \varepsilon_{1}\ldots \varepsilon_{N}}\,,
\end{equation*}
in which $\mathcal{J}^\mu (\tau)$ is an effective (operator valued) source 
\be
    \mathcal{J}^\mu (\tau) := a^\mu(x^\LCp + z^\LCp \tau/T + \xi)\frac{\ud}{\ud\tau}+\chi(\tau)n^\mu +i\sum_{i=1}^{N}\big(ik_{i}^\mu - \varepsilon_{i}^\mu\frac{\ud}{\ud\tau}\big)\delta(\tau-\tau_{i}) \;,
\ee
Since the fluctuation integral is now Gaussian, it is easily computed in terms of the worldline Green's function $\Delta(\tau_i,\tau_j)$, that is the inverse of $2\ud^2/\ud\tau^2$ with Dirichlet BCs, which is found to be
\begin{equation}\label{eq:Delta}
    \Delta_{ij} \coloneqq \Delta(\tau_{i},\tau_{j}) = \frac{1}{2}|\tau_{i}-\tau_{j}|-\frac{1}{2}(\tau_{i}+\tau_{j})+\frac{\tau_{i}\tau_{j}}{T}\,.
\end{equation}
It is easily checked that Dirichlet BCs hold: $\Delta(0,\tau_{i})=\Delta(T,\tau_{i})=\Delta(\tau_{j},0)=\Delta(\tau_{j},T)=0$. With this, the fluctuation integral becomes 
\be
\label{fluct-done}
\int_{q(0) = 0}^{q(T) = 0} \hspace{-1em} \mathcal{D}q(\tau)\,
 \e^{i\int_{0}^{T}\ud\tau[-\frac{\dot{q}^{2}}{4}-\mathcal{J}\cdot q]} = {-i(4\pi T)^{-2}}\,
 \exp\bigg[
 -i\int_{0}^{T}\ud \tau_i \ud \tau_j\,  \mathcal{J}_\mu(\tau_i)\Delta_{ij} \mathcal{J}^\mu(\tau_j)
 \bigg] \;.
 \ee
This defines the fundamental contraction for the fluctuation variable, 
\begin{equation}
    \langle q^{\mu}(\tau) q^{\nu}(\tau')\rangle = 2 i{\eta^{\mu\nu}} \Delta(\tau, \tau')\,,
\end{equation} 
and the free path integral normalisation is recovered by setting $\mathcal{J}=0$. To proceed, we wish to write out the exponent in (\ref{fluct-done}) explicitly. Note, though, that $\Delta_{ij}$ is not proper time-translation invariant due to the boundary conditions~\cite{Ahmadiniaz:2020wlm}, hence left and right proper time-derivatives must be distinguished. We denote these as follows:
\be
    \ddel_{ij}\coloneqq \frac{\ud}{\ud \tau_i} \Delta_{ij} \;,
    \qquad 
    \deld_{ij}\coloneqq \frac{\ud}{\ud \tau_j} \Delta_{ij} \;,
    \qquad
    \dddel_{ij}\coloneqq \frac{\ud^2}{\ud \tau_i^2} \Delta_{ij} \;,
    \qquad 
    \mathrm{etc.} 
\ee
With this, we write out the exponent of (\ref{fluct-done}), using that the background is transverse and on-shell ($n\cdot a=0$ and $n^2=0$) to simplify. We find, writing $a_i \equiv a(x^\LCp + z^\LCp \tau_i/\tau + \xi(\tau_i))$,
\begin{align}
\int \mathcal{J}\cdot \Delta \cdot\mathcal{J} & =
\int\!\ud \tau_i \ud\tau_j \,a_i\cdot a_j \ddeld_{ij}
+
2i\sum_{j=1}^{N}\int\!\ud\tau_i\, (\ddeld_{ij} \, a_{i}\cdot\varepsilon_{j}+i\ddel_{ij}a_{i}\cdot k_{j})\notag\\
 &+ 2i\sum_{j=1}^{N}\int\!\ud\tau_i\, \,\chi_i [\deld_{ij} \varepsilon_{j}^{\LCp} + i\Delta_{ij}k_{j}^{\LCp}] 
 -\sum_{i,j=1}^{N}[\ddeld_{ij} \varepsilon_{i}\cdot\varepsilon_{j}+2i\ddel_{ij}\varepsilon_{i}\cdot k_{j}-\Delta_{ij}k_{i}\cdot k_{j}]\,.
 \label{eq:JDelJ}
\end{align}
The trivial dependence on $\chi$ means that this field can now be integrated out, yielding a $\delta$-functional:
\begin{equation}
\int\mathcal{D}\xi\mathcal{D}\chi \, \e^{i\int d\tau\,\chi\big[\xi-2i\sum_{j=1}^{N}(\deld_{\tau \tau_{j}}\varepsilon_{j}^{\LCp}+i\Delta_{\tau\tau_{j}}k_{j}^{\LCp})\big]}
=\int\mathcal{D}\xi\, \delta\big[\xi(\tau)-2\sum_{j=1}^{N}(i\deld_{\tau \tau_{j}} \varepsilon_{j}^{\LCp}-\Delta_{\tau\tau_{j}}k_{j}^{\LCp})\big]\,.
\label{eq:aux}
\end{equation}
This $\delta$-functional has the effect of shifting the argument of the background field, such that from here on we have
\begin{equation}
\label{eq:q_drop}
a^\mu_i\equiv a^{\mu}(\tau_{i}) \equiv a^\mu\Bigl( x^{\LCp} + z^{\LCp}\frac{\tau_i}{T} +2\sum_{j=1}^{N}\big[-\Delta_{ij}k_{j}^{\LCp}+i\deld_{ij}\varepsilon_{j}^{\LCp}\big]  \Bigr)\,.
\end{equation}
The dynamical fluctuation is thus replaced by a coupling of the plane wave to the $N$ scattering photons~\cite{AntonPlane,Edwards:2021vhg,*Edwards:2021uif}. This is particular to plane wave backgrounds because (a) for $n^{2} \neq 0$ equation (\ref{eq:JDelJ}) picks up a contribution quadratic in $\chi$, whilst (b) for $n \cdot a \neq 0$ there is an additional term linear in $\chi$ that depends on the background; instead of (\ref{eq:q_drop}) one would have obtained via (\ref{eq:aux}) only an implicit equation for $a^\mu$.

All remaining background-dependent terms in (\ref{eq:aux}) may be expressed in terms of just two worldline structures, namely the worldline average and the periodic integral
\be
    \llangle f\rrangle\coloneqq T^{-1}\int_{0}^{T}\!\ud\tau \, f(\tau) \;, \qquad 
    I_{\mu}(\tau)\coloneqq\int_{0}^{\tau}\!\ud\tau' \big[ a_{\mu}(\tau')-\llangle a_{\mu}\rrangle\big] \;,
\ee
respectively. These would have to be computed for a given background once the functional form of $a_{\mu}$ has been fixed.
At this stage the path integral has (at least formally) been computed. Gathering everything together we obtain our Master Formulae for the $N$-photon
dressed  propagator
\begin{equation}\label{Master:sQED:correlator-pre-gauge}
\begin{split}
\mathcal{D}_{N}^{x'x}=
{i}(-e)^{N}
\int_{0}^{\infty}\!\ud T \, &(4i\pi T)^{-2} \e^{-i\frac{z^{2}}{4T}} {\prod_{i=1}^{N}
\int_{0}^{T} \!\ud\tau_{i}}\\
&\e^{-iM^{2}(a)T}
\Pn^{x'x}(\varepsilon_{1}, \ldots \, \varepsilon_{N}) \,
\e^{-iz\cdot\llangle a\rrangle+ i\sum_{j=1}^{N}(x+\frac{z}{T}\tau_{j}-2I(\tau_j))\cdot k_{j}-i\sum_{i,j=1}^{N}\Delta_{ij}k_{i}\cdot k_{j}}{\Big|_{\mathrm{lin. }\; \varepsilon_{1}\ldots \varepsilon_{N}}}\;,
\end{split}
\end{equation}
in which  $M^{2}(a)\coloneqq m^{2}-\llangle a^{2}\rrangle+\llangle a\rrangle^{2}$~is analogous to the Kibble `mass'~\cite{Kibble:1975vz} which typically appears in pulsed plane waves~\cite{Harvey:2012ie}, while $\Pn^{x'x}$
is defined by
\begin{equation}
\Pn^{x'x}(\varepsilon_{1}, \ldots \, \varepsilon_{N})\coloneqq  
 i^{N}e^{\sum_{i=1}^{N}\varepsilon_{i}\cdot\frac{z}{T}+2\sum_{i=1}^{N}(\llangle a\rrangle-a_{i})\cdot\varepsilon_{i}+i\sum_{i,j=1}^{N}[2i\ddel_{ij}\varepsilon_{i}\cdot k_{j} + \varepsilon_{i}\cdot\varepsilon_{k}\ddeld_{ij}]}\,.
\end{equation}
We emphasise that this Master Formula holds for any multiplicity $N\geq 0$; it would be extremely challenging to obtain this starting from the Feynman rules. Evaluating in specific cases we can check against the literature; for $N=0$, for example, we recover a one-parameter (proper-time) representation of the scalar Volkov propagator:
\begin{equation}
{\mathcal{D}_0^{x'x}}={i} \e^{-iz\cdot\llangle a\rrangle}
\int_{0}^{\infty}\!
\ud T \, (4i\pi T)^{-2} \e^{-iM^{2}(a)T}e^{-i\frac{z^{2}}{4T}} \,.
    \label{eqVolkovD}
\end{equation}
Observe that in this case $a_{\mu}(\tau) \equiv a_{\mu}\big(x^{+} + z^{+} \frac{\tau}{T} \big)$ so that, changing variables to $u = \frac{\tau}{T}$, the worldline average  becomes $T$--independent and can be taken outside the $T$-integral. It may be written as a \emph{spacetime} average (see \cite{AntonPlane}),
\begin{equation}
   \llangle a_{\mu} \rrangle =  \int_{0}^{1}\!\ud u \, a_{\mu}\big(x^{\LCp} +  z^{\LCp}u\big) = \frac{1}{x^{\prime+} - x^{\LCp}}\int_{x^{\LCp}}^{x^{\prime +}}\!\ud y \, a_{\mu}(y) \equiv \langle a_{\mu} \rangle\,,
\end{equation}
and as such $M^{2}(a) = m^{2} - \langle a^{2} \rangle + \langle a\rangle^{2}$ now corresponds exactly to the Kibble mass.

Equation (\ref{eqVolkovD}) is equivalent to the standard momentum-integral representation of the Volkov propagator, and offers a concise version of the position space propagator in~\cite{PhysRev.82.664, Borghardt1998}. For $N=1$ we recover the (2-scalar 1-photon) three point function, and so on. Since the correlators themselves are not of immediate interest, we will present these checks later, implicitly, as part of our checks on the corresponding formula for \emph{scattering amplitudes}.

The actual computation of the dressed propagator (and, later, the amplitudes) is greatly simplified by observing that we can choose the gauge $n\cdot\varepsilon = \varepsilon^\LCp = 0$. This removes the polarisation vectors from the argument of $a_\mu$, and thus extraction of the multi-linear piece of (\ref{Master:sQED:correlator}) reduces to the expansion of $\Pn(\varepsilon_{1}, \ldots \, \varepsilon_{N})$ alone. We adopt this gauge from here on in order to present the simplest possible expressions and also match to the strong field QED literature, where this gauge is common. Doing so, then, we can write the Master Formula in this gauge as
\begin{equation}\label{Master:sQED:correlator}
\mathcal{D}_{N}^{x'x}={i}
{(-e)^{N}}
\int_{0}^{\infty}\!
\ud T \, (4i\pi T)^{-2} \e^{-i\frac{z^{2}}{4T}}
{\prod_{i=1}^{N}
\int_{0}^{T} \ud\tau_{i}}\, 
\e^{-iM^{2}(a)T}
\Pn_{N}^{x'x}\e^{-iz\cdot\llangle a\rrangle+ i\sum_{i=1}^{N}(x+\frac{z}{T}\tau_{i}-2I(\tau_i))\cdot k_{i}-i\sum_{i,j=1}^{N}\Delta_{ij}k_{i}\cdot k_{j}}\;,
\end{equation}
where the polynomial $\bar{\mathfrak{P}}^{x'x}_{N}$ is defined by the expansion of the polarisation-dependent terms to multi-linear order:
\begin{equation}\label{Master:sQED:correlator2}
{\bar{\mathfrak{P}}_{N}^{x'x}}\coloneqq  
i^{N}  \e^{\sum_{i=1}^{N}\varepsilon_{i}\cdot\frac{z}{T}+2\sum_{i=1}^{N}(\llangle a\rrangle-a_{i})\cdot\varepsilon_{i}+i\sum_{i,j=1}^{N}[2i\ddel_{ij}\varepsilon_{i}\cdot k_{j} + \ddeld_{ij}\varepsilon_{i}\cdot\varepsilon_{j}]}{\Big|_{\mathrm{lin. }\; \varepsilon_{1}\ldots \varepsilon_{N}}\,.}
\end{equation}
These polynomials generalise those defined for closed worldlines in vacuum ($P_{N}$) in \cite{ChrisRev}, for open lines in vacuum ($\bar{P}_{N}$) in \cite{Schubert:2000yt}, and for the closed loop in a background field ($\mathfrak{P}_{N}$) in \cite{Edwards:2021uif} (in position space for the time being). For convenience let us  write out the first few terms:
\begin{align}
    \bar{\mathfrak{P}}_{0}^{x'x}&=1\,,\\
    \bar{\mathfrak{P}}_{1}^{x'x}&=i\Bigl[\frac{z}{T}+2(\llangle a\rrangle-a_{1})-2\lDot\Delta_{11}k_{1}\Bigr]\cdot\varepsilon_{1}\,,\\
    \bar{\mathfrak{P}}_{2}^{x'x}&=-\Bigl[\frac{z}{T}+2(\llangle a\rrangle-a_{1})-2\lDot\Delta_{11}k_{1}-2\lDot\Delta_{12}k_{2}\Bigr]\cdot\varepsilon_{1}\notag\\
    &\hphantom{=}\,\times\Bigl[\frac{z}{T}+2(\llangle a\rrangle-a_{2})-2\lDot\Delta_{21}k_{1}-2\lDot\Delta_{22}k_{2}\Bigr]\cdot\varepsilon_{2}-2i\lDot\Delta_{12}\rDot\varepsilon_{1}\cdot\varepsilon_{2}\,.
\end{align}

\subsection{Spinor QED}
We now turn to the computation of the analogous $N$-photon {dressed propagators} in spinor QED, denoting these {by $\mathcal{S}^{x'x}_N$}. Due to the spin degrees of freedom this is a Dirac matrix-valued function, but we suppress the corresponding indices for brevity. {Referring the reader to~\cite{Fradkin:1991ci,Ahmadiniaz:2020wlm} for details, we begin by writing down the analogue of the `propagator'  (\ref{propagator_pos_pi}) in an arbitrary background, but now accounting for the spin of the fermion:}

\begin{align}\label{chaos-0}
    \mathcal{S}^{x'x} &= \big(-i\slashed{D}_{x'} - m\big)\mathcal{K}^{x'x}(a)\\
    \mathcal{K}^{x'x}(a) &=
    \int_{0}^{\infty} \!\ud T\, \e^{-im^2 T} \!
    \int_{x(0) = x}^{x(T) = x'}\hspace{-1.5em}\mathcal{D}x(\tau) \, \e^{i S_{\mathrm{WL}}[x(\tau), A]}\,
    {2^{-\frac{D}{2}}}
    \mathrm{symb}^{-1}
    \oint_{\textrm{A/P}}\hspace{-0.75em}\mathcal{D}\psi(\tau) \,    
    \e^{i\widetilde{S}_{\mathrm{WL}}[\psi(\tau), x(\tau), A]}\\
    \widetilde{S}_{\mathrm{WL}}[\psi(\tau), x(\tau), A] &= \int_{0}^{T}\ud \tau \Big[\frac{i}{2}\psi\cdot\dot{\psi}
    +
    ie\big(\psi(\tau) + \eta\big) \cdot F(x(\tau)) \cdot \big(\psi(\tau) + \eta\big)\Big]\,.
\end{align}
{The kernel $\mathcal{K}^{x'x}$ contains an integral over relativistic particle trajectories, as for the scalar case, and also a path integral over Grassmann--valued fields $\psi(\tau)$, obeying anti-periodic (A/P) BCs $\psi(0)=-\psi(T)$. These represent the spin degrees of freedom of the fermion and are minimally coupled to $A$ through its field strength $F(x(\tau))$ appearing in the action ${\tilde S}_{\text{WL}}$}. An additional Grassmann variable $\eta$ also appears; the Dirac matrix structure of the propagator is produced by acting on this variable by the (inverse of) the \emph{symbolic map}, defined by
%
%
\begin{equation}
{\mathrm{symb}\Bigl\{\gamma^{[\mu_1}\gamma^{\mu_2}...\gamma^{\mu_n]}\Bigr\}=(-i\sqrt{2})^{n}\eta^{\mu_1}\eta^{\mu_2}\ldots\eta^{\mu_n}\,.}
\end{equation}
{This map converts between antisymmetric combinations of Dirac matrices (a combinatorial factor of $1/n!$ factor is assumed) and products of Grassmann variables $\eta$. Use of the symbol map avoids lengthy Dirac-matrix algebra as it automatically produces the kernel in the (even sub-algebra of the) Clifford basis of the Dirac algebra. Note that all $\eta$-dependence in (\ref{chaos}) or any of our expressions vanishes after evaluation of the inverse map; it is therefore pragmatic to state once and for all the results relevant to us in (3+1)-dimensions as}
\begin{align}
\label{symbolic_map}
    &\mathrm{symb}^{-1}\bigl\{1\bigr\}	=\mathbb{I}_{4}\,
    \qquad
    \mathrm{symb}^{-1}\bigl\{ \eta^{\mu}\eta^{\nu}\bigr\}
    =-\frac{1}{2}\gamma^{[\mu}\gamma^{\nu]}
    =-\frac{1}{4}[\gamma^{\mu},\gamma^{\nu}]\notag \\
    &\mathrm{symb}^{-1}\bigl\{ \eta^{\mu}\eta^{\nu}\eta^{\alpha}\eta^{\beta}\bigr\}
=\frac{1}{4!}\Bigl[\{\gamma^{[\mu}\gamma^{\nu]},\gamma^{[\alpha}\gamma^{\beta]}\}-\{\gamma^{[\mu}\gamma^{\alpha]},\gamma^{[\nu}\gamma^{\beta]}\}+\{\gamma^{[\mu}\gamma^{\beta]},\gamma^{[\nu}\gamma^{\alpha]}\}\Bigr]
=i\gamma_5\epsilon^{\mu\nu\alpha\beta}\,.    
\end{align}
{Now, taking $A$ as in (\ref{Total-A2}) to introduce both our background plane wave and the $N$ external photons,
we expand (\ref{chaos-0}) to multi-linear order in the photon polarisations to obtain the $N$-photon dressed propagator}
\begin{align}
\label{chaos}
    \mathcal{S}^{x'x}_N &=  
    (-i\slashed{\partial}_{x'}+\slashed{a}(x^{\prime \LCp})-m)    {\mathcal K}_{N}^{x'x}(a)+e\slashed{A}^\gamma(x') {\mathcal K}_{N-1}^{x'x}(a)\,, \\
    {\mathcal K}_{N}^{x'x}(a) &=
    (-ie)^{N}\int_{0}^{\infty} \!\ud T\, \e^{-im^2 T}\!
    \int_{x(0) = x}^{x(T) = x'}\hspace{-1.5em}\mathcal{D}x(\tau) \, \e^{i S_{\mathrm{B}}[x(\tau), a]}\,
    2^{-\frac{D}{2}}
    \mathrm{symb}^{-1}
    \oint_{\textrm{A/P}}\hspace{-0.75em}\mathcal{D}\psi(\tau) \,
    \e^{i\widetilde{S}_{\mathrm{B}}[\psi(\tau), x(\tau), a]}\prod_{i=1}^{N}V_{\eta}^{x'x}[\varepsilon_i,k_i]\,, \nonumber
\end{align}
{where $\widetilde{S}_{\mathrm{B}}[\psi(\tau), x(\tau), a]$ is given by replacing $eF(x(\tau))$ in $\widetilde{S}_{\mathrm{WL}}[\psi(\tau), x(\tau), A]$ with $f(x(\tau))$. In the `$N$-photon kernel' $\mathcal{K}_{N}^{x'x}(a)$, the proper time and bosonic integrals are the same as in the scalar case -- these represent the orbital degrees of freedom which remain unchanged. In the so-called subleading term involving ${\mathcal K}_{N-1}^{x'x}$, for each term in the sum in $\slashed{A}^{\gamma}(x')$ we remove the corresponding photon from the kernel to maintain the projection onto the multi-linear sector.} Finally, writing $\tilde{f}_{i\mu\nu}=k_{i\mu}\varepsilon_{i\nu}-k_{i\nu}\varepsilon_{i\mu}$ for the linearised field strength associated with the $i^\text{th}$ photon, the vertex operator is now given by
\begin{equation}
\label{def:spin-vertex}
    V_{\eta}^{x'x}[\varepsilon_i,k_i]
    \coloneqq
    \int_{0}^{T}\ud\tau\, \Big[\varepsilon_{i}\cdot\dot{x}(\tau_{i})
    +
    \big(\psi(\tau_{i}) + \eta\big)\cdot\tilde{f}_{i}\cdot\big(\psi(\tau_{i}) + \eta\big)\Big]
    \e^{ik_{i}\cdot x(\tau_{i})}\,,
\end{equation}
in which the second term represents the spin coupling of the external photons to the particle trajectories. 

Despite the obvious added complexity from the spin coupling to the photon fields, we stress that the same hidden Gaussianity is present here as in the scalar case. Consider again the path integral over $x^\mu$; we treat it as we did above, introducing auxiliary fields to yield a Gaussian path integral in the fluctuation $q^\mu$. While there is now an additional dependence on the background $f_{\mu\nu}$ introduced by the spin factor, this behaves in the same way as above when integrating out the auxiliary fields, i.e.~$f$ in the spin factor is ultimately evaluated at a shifted argument,
\begin{equation}
f^{\mu\nu}_i \equiv f^{\mu\nu}(\tau_{i}) \equiv
f^{\mu\nu}\Bigl(x^{\LCp}+z^{\LCp}\frac{\tau_{i}}{T}-2\sum_{j=1}^{N}\Delta_{ij}k_{j}^{\LCp}\Bigr)\,,
\end{equation}
just for $a_\mu$ earlier (recall we have gauged $\varepsilon^\LCp_i=0$ for conveniences). In short, and as is natural, the only real difference compared to the scalar case lies in the evaluation of the Grassmann path integral, which is the focus of the remainder of this section.

Observe that the vertex operators (\ref{def:spin-vertex}) introduce factors of $\psi_\eta(\tau) \equiv (\psi(\tau) + \eta)$  under the Grassmann  integral. { This motivates us to introduce the following functions,
\begin{align}
\mathfrak{W}_{\eta}(\tilde{f}_{i_{1}}; &\ldots ;\tilde{f}_{i_{S}}) 
 \coloneqq \Big\langle \psi_{\eta}(\tau_{i_{1}}) \cdot \tilde{f}_{i_{1}} \cdot \psi_{\eta}(\tau_{i_{1}}) \ldots  \psi_{\eta}(\tau_{i_{S}}) \cdot \tilde{f}_{i_{S}} \cdot \psi_{\eta}(\tau_{i_{S}})\Big\rangle\\
\hspace{-1.5em}&= 2^{-\frac{D}{2}}\oint_{\mathrm{A/P}}\mathcal{D}\psi(\tau)\, \psi_{\eta}(\tau_{i_{1}}) \cdot \tilde{f}_{i_{1}} \cdot \psi_{\eta}(\tau_{i_{1}}) \ldots  \psi_{\eta}(\tau_{i_{S}}) \cdot \tilde{f}_{i_{S}} \cdot \psi_{\eta}(\tau_{i_{S}}) \, \e^{i\int_{0}^{T}\ud \tau\, \big[\frac{i}{2}\psi\cdot\dot{\psi}+i\psi_{\eta}(\tau)\cdot f(\tau)\cdot\psi_{\eta}(\tau)\big]}\,,
\label{eq:W_0J}
\end{align}
which generalise the expectation values of the spin-part of the vertex operator introduced in vacuum ($W(\tilde{f}_{i_{1}}; \ldots ;\tilde{f}_{i_{S}})$ on the loop in \cite{ChrisRev} and $W_{\eta}(\tilde{f}_{i_{1}}; \ldots ;\tilde{f}_{i_{S}})$ for open lines in \cite{Ahmadiniaz:2020wlm}) and for one-loop amplitudes in the plane wave background ($\mathfrak{W}(\tilde{f}_{i_{1}}; \ldots ;\tilde{f}_{i_{S}})$ in \cite{Edwards:2021vhg}).} We generate the insertions under the path integral by derivatives w.r.t. a fictitious Grassmann source $\j$ (anti-commuting with $\psi$ and $\eta$),
from which follows
\begin{equation}
\mathfrak{W}_{\eta}(\tilde{f}_{i_{1}}; \ldots ;\tilde{f}_{i_{S}}) = 
\frac{\delta}{\delta \j_{i_{1}}}\cdot\tilde{f}_{i_{1}}\cdot\frac{\delta}{\delta \j_{i_{1}}} \cdots
\frac{\delta}{\delta \j_{i_{S}}}\cdot\tilde{f}_{i_{S}}\cdot\frac{\delta}{\delta \j_{i_{S}}}
\, 2^{-\frac{D}{2}}\oint_{\mathrm{A/P}}\!\mathcal{D}\psi(\tau)\, 
\e^{i\int_{0}^{T}\ud \tau\, [\frac{i}{2}\psi\cdot\dot{\psi}+i\psi_{\eta}\cdot f\cdot\psi_{\eta}+i\j\cdot\psi_{\eta}]}\bigg|_{\j=0}\,,
\label{eq:W_1J}
\end{equation}
{and the corresponding spin factor is produced through}
\begin{equation}
\mathrm{Spin}(\tilde{f}_{i_{1}}; \ldots ;\tilde{f}_{i_{S}}) \,
{\coloneqq} \,\,\mathrm{symb}^{-1}\,
\mathfrak{W}_{\eta}(\tilde{f}_{i_{1}}; \ldots ;\tilde{f}_{i_{S}})\,.
\label{eq:W_1}
\end{equation}
To compute the integral in (\ref{eq:W_1J}) we require the (spinor) worldline propagator in the field, $\mathfrak{G}^{\mu\nu}{(\tau, \tau')}$. This will define the fundamental contraction between the Grassmann fields,
\begin{equation}
    \big\langle \psi^{\mu}(\tau) \psi^{\nu}(\tau')\big\rangle = \frac{1}{2}\mathfrak{G}^{\mu\nu}(\tau, \tau')\,.
\end{equation} 
From the quadratic part of the operator appearing in the path integral action, $\mathfrak{G}$ must obey
\be\label{to-solve-cal-G}
    \Big(\frac{1}{2}\eta_{\mu\sigma}\frac{\ud}{\ud\tau}+f_{\mu\sigma}{(\tau)}\Big)\mathfrak{G}^{\sigma\nu}{(\tau,\tau')}
    =
    \eta_{\mu}{}^{\nu}\,\delta(\tau-\tau') \;,
\ee
as well as anti-periodic boundary conditions $\mathfrak{G}(0,\tau')=-\mathfrak{G}(T,\tau')$ and $\mathfrak{G}(\tau,0)=-\mathfrak{G}(\tau, T)$.
Observe that $\mathfrak{G}$ has the anti-symmetric property $\mathfrak{G}^{\mu\nu}(\tau,\tau')=-\mathfrak{G}^{\nu\mu}(\tau',\tau)$. The general {homogeneous} solution of (\ref{to-solve-cal-G}) for \emph{arbitrary} $f(\tau)$ is written conveniently in terms of an auxiliary function $\mathcal{O}(\tau, \tau')$, which takes care of the ordering of $\tau$ and $\tau'$, defined~by 
\be
   \mathcal{O}(\tau,\tau')=\mathcal{P}^{\star}\,e^{-2\int_{\tau'}^{\tau}d\sigma f(\sigma)}\,,
\ee
where 
$\Theta$ is the Heaviside step function, $\mathcal{P}^{\star} \equiv \mathcal{P}^{\star}(\tau, \tau') = \Theta(\tau-\tau') \mathcal{P} + \Theta(\tau' - \tau)\bar{\mathcal{P}}$ with $\mathcal{P}$ ($\bar{\mathcal{P}}$) denoting (anti-)path ordering in proper time and {we have made use of a matrix form for the Lorentz indices (with respect to which $\mathcal{O}$ is orthogonal)}. With the homogeneous solution, we can then find the general solution to (\ref{to-solve-cal-G}) with appropriate anti-periodic boundary conditions as
\be
    \mathfrak{G}(\tau,\tau')= \mathrm{sgn}(\tau-\tau')\mathcal{O}(\tau,\tau') + \mathcal{O}(\tau,0){\frac{1-\mathcal{O}(T,0)}{1+\mathcal{O}(T,0)}}\mathcal{O}(0,\tau')\,,
\ee
However, there are notable simplifications in our particular case that $f$ is a plane wave because, as is well known, the field strength is then nilpotent of order three. Further, $f$ evaluated at different $\tau$ commute. The Green function thus reduces to\footnote{{This is an alternative way of writing the Green function given in equation (45) of~\cite{Edwards:2021vhg}, with the advantage of being manifestly gauge invariant. There $\mathfrak{G}^{\mu\nu}$ was written in terms of periodic integrals of the derivative of $a(\tau)$ which made its anti-periodicity easier to see.}}
\begin{align}
\mathfrak{G}(\tau,\tau')&= \e^{-2 \int_{\tau'}^{\tau}d\sigma\, f(\sigma)} \Big[ \mathrm{sgn}(\tau - \tau') + \tanh\Big( \int_{0}^{T}d\sigma\, f(\sigma) \Big) \Big] \\
&=\mathrm{sgn}(\tau-\tau')
\Bigl[1-2\int_{\tau'}^{\tau}\!\ud\sigma \,f(\sigma)
+2\Big(\int_{\tau'}^{\tau}\!\ud \sigma\, f(\sigma)\Big)^{2}\Bigr]
+T\llangle f\rrangle\Bigl[1-2\int_{\tau'}^{\tau}\!\ud \sigma\, f(\sigma)\Bigr]\,.
\end{align}
Equipped with the Green function, we compute the integral in (\ref{eq:W_1J}) by completing the square, using the shift \newline $\tilde{\psi}(\tau)=\psi(\tau)+\int\!\ud\tau'\mathfrak{G}(\tau,\tau')\cdot(f(\tau')\cdot \eta+\frac{1}{2}{\theta}(\tau'))$ The integral over $\tilde\psi$ then generates the determinant $\mathrm{Det}(\frac{1}{2}\frac{\ud}{\ud\tau}+f)$ (for anti-periodic boundary conditions) which because of the nilpotency of $f$ simply gives a factor of $2^{\frac{D}{2}}$, being the number of degrees of freedom of the fermion in $D$ (even) space-time dimensions (this should be contrasted with the constant field case, where the normalisation picks up a non-trivial field dependence \cite{Reuter:1996zm, Shaisultanov:1995tm}).

Gathering all of the above together, the Grassmann integral as defined in (\ref{eq:W_1J}) becomes
\begin{equation}\label{generating_functional}
\mathfrak{W}_{\eta}(\tilde{f}_{i_{1}}; \ldots ;\tilde{f}_{i_{S}}) 
=
\frac{\delta}{\delta \j_{i_{1}}}\cdot\tilde{f}_{i_{1}}
\cdot\frac{\delta}{\delta \j_{i_{1}}}\cdots
\frac{\delta}{\delta \j_{i_{S}}}\cdot\tilde{f}_{i_{S}}\cdot\frac{\delta}{\delta \j_{i_{S}}}
\e^{-\int_{0}^{T}\!\ud\tau \, [\eta \cdot f(\tau) \cdot \eta +{\j(\tau)} \cdot \eta]
-\int_{0}^{T}\!\ud\tau
\ud\tau'[\eta \cdot f(\tau) \cdot \mathfrak{G}(\tau, \tau') \cdot \j(\tau')+\frac{1}{4}\j(\tau)\cdot \mathfrak{G}(\tau, \tau')\cdot \j(\tau')]}\Big|_{\j=0}\,.
\end{equation}

The Grassmann path integral is therefore formally computed. In particular,
\begin{align}
\label{W_0}    \mathfrak{W}_{\eta}(\emptyset) &= e^{-\int_{0}^{T}d\tau\, \eta\cdot f(\tau)\cdot \eta}\,, \\
\label{W_f_1}    \mathfrak{W}_{\eta}(\tilde{f}_{i_{1}}) &= \Bigl\{-\tfrac{1}{2}\mathrm{tr}[\tilde{f}(\tau_{i_{1}})\cdot\mathfrak{G}         (\tau_{i_{1}},\tau_{i_{1}})]+\eta\cdot\mathfrak{G}^{\mathbb{T}}(\tau_{i_{1}})\cdot\tilde{f}(\tau_{i_{1}})\cdot\mathfrak{G}(\tau_{i_{1}})\cdot\eta\Bigr\} e^{-\int_{0}^{T}d\tau\, \eta\cdot f(\tau)\cdot\eta} \,,\\ 
\label{W_f_2}    \mathfrak{W}_{\eta}(\tilde{f}_{i_{1}};\tilde{f}_{i_{2}}) &=
    \Biggl\{\bigl[-\tfrac{1}{2}\mathrm{tr}[\tilde{f}(\tau_{i_{2}})\cdot\mathfrak{G}(\tau_{i_{2}},\tau_{i_{2}})]+\eta\cdot\mathfrak{G}^{\mathbb{T}}(\tau_{i_{2}})\cdot\tilde{f}(\tau_{i_{2}})\cdot\mathfrak{G}(\tau_{i_{2}})\cdot\eta\bigr]\times\bigl[\tau_{i_{2}}\to\tau_{i_{1}}\bigr]\notag\\
    &\hspace{2.5em} -\tfrac{1}{2}\mathrm{tr}[\tilde{f}(\tau_{i_{1}})\cdot\mathfrak{G}(\tau_{i_{1}},\tau_{i_{2}})\cdot\tilde{f}(\tau_{i_{2}})\cdot\mathfrak{G}(\tau_{i_{2}},\tau_{i_{1}})]\notag\\
    &\hspace{2.5em} +2\eta\cdot\mathfrak{G}^{\mathbb{T}}(\tau_{i_{2}})\cdot\tilde{f}(\tau_{i_{2}})\cdot\mathfrak{G}(\tau_{i_{2}},\tau_{i_{1}})\cdot\tilde{f}(\tau_{i_{1}})\cdot\mathfrak{G}(\tau_{i_{1}})\cdot\eta\Biggr\} e^{-\int_{0}^{T}d\tau\, \eta\cdot f(\tau)\cdot\eta}\,,
\end{align}
where $\mathfrak{G}_{\mu\nu}(\tau_{i})\coloneqq\eta_{\mu\nu}-\int_{0}^{T}d\tau[\mathfrak{G}(\tau_{i},\tau)\cdot f(\tau)]_{\mu\nu}$ and $\mathbb{T}$ denotes the transpose in Lorentz indices -- in particular we have $\mathfrak{G}^{\mathbb{T}}{}_{\mu\nu}(\tau_{i}) = \eta_{\mu\nu} - \int_{0}^{T} d\tau [f(\tau) \cdot \mathfrak{G}(\tau, \tau_{i})]_{\mu\nu}$.

Putting all of this together, the $N$-photon dressed propagator can be written in a `spin-orbit decomposition' by summing over assignation of the $N$ external photons to either the spin or bosonic part of the vertex~ \cite{Edwards:2021vhg}, as follows:
\begin{align}
\label{Master:QED:correlator}
\mathcal{S}_{N}^{x'x} & = 
(-i\slashed{\partial}_{x'}+\slashed{a}(x^{\prime \LCp})-m)    {\mathcal K}_{N}^{x'x}(a)+e\slashed{A}^\gamma(x') {\mathcal K}_{N-1}^{x'x}(a)\,, \\
 \mathcal{K}_{N}^{x'x}(a) &= \sum_{S=0}^{N}\sum_{\{i_{1}:i_{S}\}}\mathcal{K}_{NS}^{\{i_{1}:i_{S}\}x'x}(a)\,,\\
\mathcal{K}_{NS}^{\{i_{1}:i_{S}\}x'x}(a) & ={i}(-e)^{N}\int_{0}^{\infty}\!\ud T\, (4\pi iT)^{-2}\, \e^{-iM^{2}(a)T-i\frac{z^{2}}{4T}-iz\cdot\llangle a\rrangle} \nonumber \\
 & \times\prod_{i=1}^{N}
 \int_{0}^{T}\!\ud \tau_{i}\,  \mathrm{Spin}(\tilde{f}_{i_{1}}; \ldots ;\tilde{f}_{i_{S}}) \,\bar{\mathfrak{P}}_{NS}^{\{i_{1}:i_{S}\}x'x}\,
 \, \e^{i\sum_{i=1}^{N}\big[x+\frac{z}{T}\tau_{i}-2I(\tau_i)\big]\cdot k_{i}-i\sum_{i,j=1}^{N}\Delta_{ij}k_{i}\cdot k_{j}}\,.
\end{align}
The sum on the second line runs over the allocation of $S$, out of the $N$, photons to the spin part of the vertex operator, $V_{\eta}^{x'x}$, which subsequently appear in $\mathrm{Spin}(\tilde{f}_{i_{1}}; \ldots ;\tilde{f}_{i_{S}})$. Then the remaining $N-S$  photons appear in the polynomial $\bar{\mathfrak{P}}_{NS}^{\{i_{1}:i_{S}\}x'x}$, defined by 

\begin{equation}
\bar{\mathfrak{P}}_{NS}^{\{i_{1}:i_{S} \}x'x}\coloneqq  i^{{N-S}} \e^{\sum_{i=1}^{N}\varepsilon_{i}\cdot\frac{z}{T}+2\sum_{i=1}^{N}\big[(\llangle a\rrangle-a_{i})\cdot\varepsilon_{i}\big]+i\sum_{i,j=1}^{N}\big[\varepsilon_{i}\cdot\varepsilon_{k}\ddeld_{ij}+2i\ddel_{ij}\varepsilon_{i}\cdot k_{j}\big]}\Big|_{\varepsilon_{i_{S+1}}...\varepsilon_{i_{N}}}^{\varepsilon_{i_{1}}...\varepsilon_{i_{S}}=0}\,.
\end{equation}
where the notation on the far right means that the polarisation vectors $\varepsilon_{i_{1}}$ to $\varepsilon_{i_{S}}$ should be put to zero before the remaining expression is expanded to multi-linear order in the $\varepsilon_{i_{S+1}}$ to $\varepsilon_{i_{N}}$. These polynomials generalise those introduced in vacuum ($\bar{P}_{NS}^{\{i_{1}; i_{S}\}}$) in \cite{Ahmadiniaz:2020wlm} and satisfy
\begin{align}
\bar{\mathfrak{P}}_{N0}^{\{ \}x'x} = \bar{\mathfrak{P}}_{N}^{x'x}\,, \qquad   \bar{\mathfrak{P}}_{NN}^{\{1:N\}x'x} = 1\,.
\end{align}
Again, these are position space expressions, but below we shall transform to momentum space for the purpose of evaluating scattering amplitudes. Although this Master Formula appears lengthy, it is important to emphasise that it represents a formal evaluation of the path integral for an arbitrary number of photons inserted along the background-dressed propagator, conveniently split into contributions from the vertex function representing orbital interactions (in $\bar{\mathfrak{P}}_{NS}^{\{i_{1}:i_{S} \}x'x}$) and spin interactions (in $\mathrm{Spin}(\tilde{f}_{i_{1}}; \ldots ;\tilde{f}_{i_{S}})$). All of these insertions are integrated along the particle trajectories, so that the Master Formula represents a sum over all Feynman diagrams contributing to the dressed propagator that differ by permutation of the external photons. Obtaining such a formula from the standard formalism (Furry picture, say) of strong field QED would be a significantly more complicated task.

For completeness, we note that the $N = 0$ case provides a worldline representation of the well-known Volkov propagator as a one-parameter integral
\begin{equation}
    \mathcal{S}^{x'x} = {i}\big(-i(\slashed{\partial}_{x'} + i\slashed{a}(x^{\prime +})) - m\big)\e^{-i z \cdot \langle a \rangle}\int_{0}^{\infty} \!\ud T\, (4\pi i T)^{-2}\e^{-i M^{2}(a)T - i \frac{z^{2}}{4T} + {\frac{T}{z^{+}}}
    \big[\slashed{n} \slashed{a}(x^{\prime +}) + \slashed{a}(x^{+}) \slashed{n}\big]}\,,
\end{equation}
where we used
$\mathrm{Spin}(\slashed{0}) = 1 + {\frac{T}{2}}\gamma \cdot \langle f \rangle \cdot \gamma =
{1 + {T} \slashed{n} \langle \slashed{a}'\rangle}$,
computed the integral {in the average explicitly} and re-exponentiated using $\slashed{n}^{2} = 0$. This is again equivalent to other representations of the Volkov propagator  \cite{PhysRev.82.664, Borghardt1998, Fedotov:2022ely}.

\section{LSZ for scattering amplitudes}
\label{sect:LSZ}
The objective of this section is to take the Master Formulae for the dressed propagators $D^{x'x}_N$ and $S_N^{x'x}$ above and produce from them equivalent Master Formulae for (2-scalar) $N$-photon scattering amplitudes (for $N \geqslant 1$). To do so we must perform LSZ reduction on the two massive, external legs of the dressed propagators.

In previous worldline literature, amputation was often done `by-hand', by obtaining the $N$-point correlation functions in momentum space and then -- once the proper-time integral had been computed -- removing external legs with the appropriate inverse matter propagators~\cite{Ahmadiniaz:2020wlm, Ahmadiniaz:2021gsd}. Only then could the external particles be taken on-shell -- the proper-time integral produces the pole structure of the correlation functions with respect to external matter legs and so is divergent in the on-shell limit. This is a notable example where the Feynman diagram prescription to omit external propagators had appeared less trivial from a worldline perspective.
Recently, however, \cite{Bonocore:2020xuj, Mogull:2020sak} showed how amputation can be achieved under the proper-time integral for scalar matter legs, with the inverse propagators simply modifying the bounds on the proper-time and parameter integrals.
This exposes the on-shell residue of the correlation functions without the need to carry out amputation by hand. We will here generalise this approach to spinor theories, and also show it is unspoiled by the plane-wave background.

To perform LSZ we draw the external legs out to asymptotic times and Fourier transform. Alternatively, we can Fourier transform to momentum space and find the residues of the dressed propagator as the momenta are taken onto the mass-shell. Starting with scalar QED, the amplitude takes the form
\begin{align}
\label{def:ampliude-N_pos}
\mathcal{A}_{N}^{p'p}&= {-\lim_{p'^2,p^2\rightarrow m^2}}\int\!\ud^4x'\ud^4x\, \e^{i(p'+ a^\infty)\cdot x'-ip \cdot x}\, 
    {\left[(\partial_{x'}+ia^\infty)^2+m^2\right]}
    {\big[\partial^2_{x}+m^2\big]}
    \,\mathcal{D}_{N}^{x'x}\,,\\
    \label{def:ampliude-N}
    &=\lim_{p'^2,p^2\rightarrow m^2}-(p'^2-m^2)(p^2-m^2)\mathcal{D}_{N}^{\tilde{p}'p}\,,
\end{align}
where in the second line we defined $\tilde{p}' = p + a^{\infty}$ and introduced the momentum-space propagator $\mathcal{D}_N^{p'p}$, defined by
\begin{align} \label{propagator_mom_def}
    \mathcal{D}_N^{p'p} &:=\int\!\ud^4x'\ud^4 x\, \e^{ip'\cdot x'-ip \cdot x}\, \mathcal{D}^{x'x}_N \;.
\end{align}
The expression~\eqref{def:ampliude-N_pos} is (almost) textbook-standard LSZ in position space but to compensate for the fact that our potential becomes pure gauge in the far future, the on-shell, outgoing momentum $p'$ in the Fourier kernel is shifted to ${\tilde p}' = p' + a^\infty$~\cite{Kibble:1975vz,Dinu:2012tj}. The expression \eqref{def:ampliude-N} makes it clear that the amplitude $\mathcal{A}_{N}^{p'p}$ is the residue of $\mathcal{D}^{\tilde{p}'p}_N$ at on-shell momenta. {In our conventions $\mathcal{A}_{N}^{p'p}$ describes $N$-photon \emph{emission} from a particle traversing the plane wave. Absorption and pair-production/annihilation amplitudes are of course obtained by crossing.}

Similarly for the spinor case, starting from the master formula for the dressed propagator (\ref{Master:QED:correlator}), we can extract the spin-polarised amplitude $\mathcal{M}_{Ns's}^{p'p}$ as
\be
\label{QED:amplitude:start}
   \mathcal{M}_{Ns's}^{p'p} = i{\lim_{p'^2,p^2\rightarrow m^2}} \int \!\ud^{4}x'\ud^{4}x\,
    \e^{i\tilde{p}'\cdot x'-ip\cdot x}
    \bar{u}_{s'}(p')(i\slashed{\partial}_{x'}-\slashed{a}^{\infty}-m)\mathcal{S}_{N}^{x'x}(-i\overleftarrow{\slashed{\partial}}_{x}-m)u_{s}(p)\,,
\ee
in which $\bar{u}_{s'}(p')$ and $u_{s}(p)$ are free Dirac spinors. We now proceed to perform the LSZ reduction explicitly, starting with scalar QED.

\subsection{Scalar QED}
We begin by evaluating the momentum-space propagator via direct Fourier transform of the Master Formula (\ref{Master:sQED:correlator}):
\be
    \mathcal{D}_N^{\tilde{p}'p} =\int\!\ud^4x'\ud^4 x\, \e^{i{\tilde p}'\cdot x'-ip \cdot x}\, \mathcal{D}^{x'x}_N \;.
\ee
The integrals over $x'^{\LCperp,\LCm}$ and $x^{\LCperp,\LCm}$ generate\footnote{{To evaluate similar integrals in the existing literature it was found to be convenient to change variables to endpoint centre of mass and relative separation ($z$). However, for our later LSZ amputation of the external legs it is more useful to integrate separately with respect to the endpoint coordinates.}}, as in the vacuum case, four $\delta$-functions, explicitly ${\delta^3_{\perp,\LCm} \big(\tilde{p}'+K-p\big)}$$\delta(x^{+}-x'^{+}+2g^{+}+2p'^{+}T)$, where we write $K=\sum_{i=1}^{N} k_i$ to compactify notation. The first three $\delta$-functions describe the (expected) conservation of lightfront three-momentum in the plane wave background. The final $\delta$-function allows us to trivially perform e.g.~the $x'^{\LCp}$ integral, so that we can replace $x'^{+}\rightarrow x^{+}+2g^++2p'^+T$ in what remains; in particular, the classical trajectory on which the gauge field depends throughout $\mathcal{D}_N^{x'x}$, as in \eqref{eq:q_drop}, is modified to, where $g \equiv g(\{\tau_i\}):=\sum_{i=1}^{N}(k_i\tau_i-i\epsilon_i)$,
\begin{equation}\label{eq:x_cl}
    x^\LCp_{cl}(\tau)=x^\LCp + g^{\LCp}
    +(p'+p)^\LCp\tau-\sum_{i=1}^Nk_{i}^{\LCp}|\tau-\tau_i|\,.
\end{equation}
Thus we can do all but one of the Fourier integrals, which eventually yield
\begin{align}\label{Master:sQED:momentum-propagator}
    \mathcal{D}_N^{\tilde{p}'p} 
&=(-ie)^N(2\pi)^3{\delta^3_{\perp,\LCm} \big(\tilde{p}'+K-p\big)}\int_{0}^{\infty}\!\ud T\,  \e^{i(p'^2-m^2+i 0^{\LCp})T} \int_{-\infty}^{\infty}\ud x^\LCp \,\e^{i(p'_{\LCp}+K_{\LCp}-p_{\LCp})x^{\LCp}}\\
    \nonumber
	&\times\prod_{i=1}^{N}\int_{0}^{T}\ud\tau_i \,\ \e^{ - 2ig\cdot \llangle a \rrangle -2 i T p' \cdot \llangle \delta a \rrangle + i T \llangle \delta a^{2} \rrangle -2i\sum_{i=1}^{N}\left[k_{i} \cdot I(\tau_i)-i\varepsilon_i \cdot I'(\tau_i)\right]} \\ \nonumber
 &\hspace{5em}\times \e^{i g\cdot (2\tilde{p}'+K)-i\sum_{i,j=1}^{N}\left(\frac{|\tau_i-\tau_j|}{2}k_i\cdot k_j-i\,\textrm{sign}(\tau_i-\tau_j)\varepsilon_i \cdot k_j+\delta(\tau_i-\tau_j)\varepsilon_i \cdot \varepsilon_j\right)} \Big\vert_{\textrm{lin. }\varepsilon_{1} \ldots \varepsilon_{N}}\,,
\end{align}
{in which we have defined $\delta a(x^{+}):=a(x^{+})-a^{\infty}$ and $a(\tau)\equiv a(x^{\LCp}_{\rm cl}(\tau))$.} Note that in the vacuum limit $a_{\mu} \rightarrow 0$ we can carry out the $\hat{x}^{+}$ integral to complete the conservation of $4$-momentum and so recover one version of the Master Formula given in \cite{Ahmadiniaz:2020wlm} and \cite{Shaisultanov:1995tm}.

{To convert \eqref{Master:sQED:momentum-propagator}} into a master formula for the amplitudes, we have to perform LSZ on each massive scalar leg (these are produced by the parameter and proper time integrals). To do so we observe that (\ref{def:ampliude-N}) has, using (\ref{Master:sQED:momentum-propagator}), the following form, writing down only the relevant structures:
\be
    -i({p'}^2-m^2+i 0^\LCp) \int_0^\infty\!\ud T\, \e^{ i ({p'}^2-m^2 + i 0^\LCp) T} F(T) \;.
    \label{eqFT}
\ee
The on-shell limit ${p^2\to m^2-i0^{+}}$ therefore returns the residue of the mass-shell pole of the function defined by the integral. To isolate this pole we proceed as in~\cite{Laenen:2008gt,Bonocore:2020xuj,Mogull:2020sak} where LSZ was considered for, e.g.,~the $N$-graviton-dressed propagator in vacuum\footnote{We note in passing that the same `trick' is useful in exposing the connection between gauge invariance and infra-red behaviour of amplitudes in background plane waves~\cite{Ilderton:2020rgk}}. We integrate by parts (off-shell) in order to expose the residue, as so:
\be\begin{split}
-i({p'}^2-m^2 + i 0^\LCp)\int_{0}^{\infty}\!\ud T\, \e^{i({p'}^2-m^2 + i 0^\LCp)T} F(T) 
=F(0) + \int_0^\infty\!\ud T \,  \e^{i({p'}^2-m^2 + i 0^\LCp)T}\frac{\ud}{\ud T} F(T) \;.
\end{split}
\ee
We can now take ${p'}^2\to m^2$ and $0^\LCp\to 0$ (in either order), upon which the integral becomes exact, and we have
\be
  \lim_{{p'}^2\to m^2}  -i({p'}^2-m^2+i 0^\LCp) \int_0^\infty\!\ud T\, \e^{ i ({p'}^2-m^2 + i 0^\LCp) T} F(T)  = F(\infty)\;.
\ee
Ultimately, then, performing the first amputation on (\ref{Master:sQED:momentum-propagator}) is equivalent to dropping the integral over proper time $T$ {and its accompanying mass-shell exponent}, and taking the limit $T\rightarrow \infty$ of what remains (this is the same argument as in vacuum, which we comment on further after performing the second amputation, below). 
We thus find
\begin{align}
 &\lim_{p'^2\rightarrow m^2}-i(p'^2-m^2+i0^\LCp) {\mathcal{D}^{p'p}_{N}}
	=(-ie)^N(2\pi)^3\delta^3_{\LCperp,\LCm} (\tilde{p}'+K-p) \int_{-\infty}^{\infty}\!\ud x^{\LCp}  \e^{ i\left(p'_\LCp+K_\LCp-p_\LCp\right)x^\LCp}
 \prod_{i=1}^{N}
 \int_{0}^{\infty}\!\ud \tau_i\,\label{AppEq:Amp1}\\\nonumber
    & \e^{-i\int_{0}^{\infty}[2p'\cdot\delta a(\tau)-\delta a^{2}(\tau)]\ud\tau -2i\sum\limits_{i=1}^{N}\left[\int_{0}^{\tau_i}k_{i}\cdot a(\tau)\ud\tau-i\varepsilon_i\cdot a(\tau_i)\right] 
    + i g \cdot (2\tilde{p}'+K)
    - i\sum\limits_{i,j=1}^{N} (\frac{|\tau_i-\tau_j|}{2}k_i\cdot k_j-i\,\mathrm{sgn}(\tau_i-\tau_j)\varepsilon_i\cdot k_j+\delta(\tau_i-\tau_j)\varepsilon_i\cdot\varepsilon_j)}\bigg|_{\textrm{lin. }\varepsilon_{i} \ldots \varepsilon_{N}}\,.
\end{align}
We note that all terms with worldline {averages have ultimately been replaced with (convergent) integrals over $\mathbb{R}^{+}$. This was the advantage of having computed the Fourier integrals with respect to the individual endpoints as discussed above. Equation (\ref{AppEq:Amp1}) is the one-side amputated propagator}. 

Turning to the amputation with respect to $p$, at this stage it is advantageous to introduce the mean and deviation proper time variables as follows:
\be
	\tau_0 \coloneqq \frac{1}{N}\sum_{i=1}^N\tau_i \;,
    \qquad
    \bar{\tau}_i\coloneqq \tau_i-\tau_0 \;.
\ee
The reason for this change of variable is that it allows us to re-express (\ref{AppEq:Amp1}) in a form which renders the \emph{second} LSZ amputation immediate. To achieve this,
we first rewrite the proper time integrals appearing in (\ref{AppEq:Amp1}) in terms of the new variables as ({note the factor of $\frac{1}{N}$ in the $\delta$-function is missing in (3.18) of \cite{Mogull:2020sak}})
\be
    \prod_{i=1}^N\int_{0}^{\infty}\!\ud\tau_i
    =\int_{-\infty}^{\infty}\!\ud\tau_0
    \prod_{i=1}^N\int_{-\infty}^{\infty}\!\ud \bar{\tau}_i \, \delta\bigg(\sum_{j=1}^N\frac{\bar{\tau}_j}{N}\bigg) \;,
    \label{eqAmpp}
\ee
We also make a change of variable for the $x^\LCp$-integration, $	\bar{x}^\LCp := {x^\LCp+(p'+p+K)^\LCp\tau_0+ g^\LCp(\{\bar{\tau}_{i}\})}$, and it is convenient to change variables in all $\ud\tau$ integrals from $\tau$ to $\bar{\tau} \coloneqq \tau-\tau_0$, such that the background gauge field now appears as
\begin{align}\label{NeumannBC}
	a(\bar\tau) \equiv a \big(
 \bar{x}^{\LCp}+(p'+p)^\LCp\,{\bar\tau}
    -\sum_{i=1}^{N}
    k_{i}^{\LCp}|{\bar\tau}-\bar{\tau}_i|
    \big) \;.
\end{align}
In terms of the shifted variables $\{\bar{x}^{\LCp},\tau_0,\bar{\tau}_i\}$, the once-amputated propagator (\ref{AppEq:Amp1}) takes the form
\begin{equation}
\label{eq:added}
 (-ie)^N(2\pi)^3\delta_{\LCperp,\LCm}(\tilde{p}'+K-p)\!
 \int_{-\infty}^{\infty}\!\ud\bar{x}^\LCp\,\e^{ i(K+p'-p)_\LCp\bar{x}^{\LCp}}
 \int_{0}^{\infty}\!\ud \tau_0 \,
 \e^{i{(p^2-m^2)}\tau_0}
 \int_{-\infty}^{\infty}\prod_{i=1}^{N}\ud\bar{\tau}_i\,
 \delta\bigg(\sum_{i=1}^N\frac{\bar{\tau}_i}{N}\bigg) G(\tau_{0}) \;,
\end{equation}
in which the function appearing in the factor is
\begin{align}
\label{eq:G-big-def}
G(\tau_{0})&=\e^{ -i(2 p'+a^{\infty})\cdot a^{\infty}\tau_0   -i\int_{-\tau_0}^{\infty}\!\ud\bar{\tau}\, [2p'\cdot \delta a(\bar{\tau})-\delta a^{2}(\bar{\tau})] 
{-2i\sum_{i=1}^{N} \left[\int_{-\tau_0}^{\bar{\tau}_i}\!\ud\bar{\tau}\, k_{i}\cdot a(\bar{\tau})-i\varepsilon_i\cdot a(\bar{\tau}_i) 
\right] }} \\\nonumber
&\times \e^{i(\tilde{p}'+p)\cdot g -i\sum_{i,j=1}^{N}\left(\frac{|\bar{\tau}_i-\bar{\tau}_j|}{2}k_i \cdot k_j-i\,\mathrm{sgn}(\bar{\tau}_i-\bar{\tau}_j)\varepsilon_i \cdot k_j+\delta(\bar{\tau}_i-\bar{\tau}_j)\varepsilon_i\cdot \varepsilon_j\right)}\;.\bigg\vert_{\textrm{lin. }\varepsilon_{1}, \ldots \varepsilon_{N}}
\end{align}
Note that the factor $-i(2 p'+a^{\infty})\cdot a^{\infty}\tau_0$ in the exponential diverges in the $\tau_0\rightarrow\infty$ limit, but can be absorbed into the Volkov-like term, also divergent in the same limit, to yield the convergent factor $-i\int_{-\tau_0}^{0}[2\tilde{p}' \cdot a(\bar{\tau})-a^{2}(\bar{\tau})]\ud\bar{\tau}-i\int_{0}^{\infty}[2p'\cdot\delta a(\bar{\tau})-\delta a^{2}(\bar{\tau})]\ud\bar{\tau}$. After this rearrangement, one finds that the dependence on $\{p^2-m^2,\tau_0\}$ {in (\ref{eq:added})--(\ref{eq:G-big-def})} exactly mirrors the dependence on $\{p^{\prime2}-m^2, T\}$ in the original expression, before the first amputation. Thus we can simply repeat the previous LSZ argument but applied to $\{p^2-m^2,\tau_0\}$ in order to extract the pole at the \emph{incoming} mass-shell; effectively this removes the integral over $\tau_0$ and takes $\tau_0\rightarrow \infty$ in the remainder, yielding our final master formula for the 2-scalar $N$-photon scattering amplitudes:
 \begin{align}\label{Master:sQED:amplitude}
 	\mathcal{A}^{p'p}_{N}&=(-ie)^N(2\pi)^3\delta_{\perp,-}({\tilde p}'+K-p)\int_{-\infty}^{\infty}\!\ud x^{\LCp}e^{i(K+p'-p)_\LCp x^{\LCp}}
 \int_{-\infty}^{\infty}\prod_{i=1}^{N} \ud\tau_i\, 
 \delta\bigg(\sum_{j=1}^N\frac{\tau_j}{N}\bigg)
  \nonumber \\
 &\times 
 \e^{-i\int_{-\infty}^{0}
 [2\tilde{p}' \cdot a(\tau)-a^{2}(\tau)]\ud\tau
 -i\int_{0}^{\infty}\!
 [2p'\cdot\delta a(\tau)-\delta a^{2}(\tau)]\ud\tau
 -2i\sum_{i=1}^{N}
 [\int_{-\infty}^{\tau_i}k_{i} \cdot a(\tau)\ud\tau-i\varepsilon_i\cdot a(\tau_i)]}\nonumber \\
 &\times e^{i({\tilde p}' +p)\cdot g-i\sum_{i,j=1}^{N}\bigl(\frac{|\tau_i-\tau_j|}{2}k_i \cdot k_j-i\,\mathrm{sgn}(\tau_i-\tau_j)\varepsilon_i \cdot k_j+\delta(\tau_i-\tau_j)\varepsilon_i \cdot \varepsilon_j\bigr)}
 \bigg|_{\textrm{lin. }\varepsilon} \;,
 \end{align}
where $a(\tau)$ is as in (\ref{NeumannBC}), and we have simply relabelled
${\bar x}^\LCp \to x^\LCp$, and $\bar{\tau},\bar{\tau}_i\to \tau,\tau_i$.

There are several features of this all-orders formula worth discussing. First, as a consistency check, it is straightforward to check that in the vacuum limit ($a\to 0$) the $x^\LCp$ integral can again be performed and one recovers the known results in~\cite{Daikouji:1995dz,Ahmad:2016vvw,Mogull:2020sak}. 
Second, similarly to~\cite{Mogull:2020sak}, a short set of rules summarises the LSZ reduction. The first three are shared with the vacuum case~\cite{Mogull:2020sak}: (i) drop the $T$ integral, (ii) insert $\delta(\sum_{j=1}^N\tau_j/N)$ and (iii) take the $\ud\tau_i$ and $\ud\tau$ integrals over $\mathbb{R}$. Here, beyond the vacuum case, there are additional rules: iv) drop all worldline averages and (v) `introduce' the divergent factor $\int_{-\infty}^{0}-2i \tilde{p}' \cdot a^\infty d\tau$ into the exponential, which ensures that the proper-time integral is convergent in the asymptotic past -- we stress that this `by hand' addition only occurs at the level of these rules, it emerges naturally as part of LSZ reduction, as described above.

Third, the change in integration range for the $\ud\tau_i$-integrals can be understood as manifesting the fact that $\mathcal{A}^{p'p}_{N}$ is an asymptotic quantity, while the purpose of $\delta(\sum_{j=1}^N \tau_j/N)$ is to `gauge' the proper-time translational symmetry of the system.
Clearly neither of these features should be particular to any choice of background that tends to at most a constant asymptotically, and indeed they are the same in our plane wave background as in vacuum.

Finally, we observe that $x_{cl}^{\LCp}(\tau)$ in (\ref{NeumannBC}) solves the classical worldline equation of motion with the boundary conditions $\tfrac{1}{4}\dot{x}^{\LCp}(-\infty)=p_{\LCm}$ and $\tfrac{1}{4}\dot{x}^{\LCp}(\infty)=p'_{\LCm}$. It is natural for this solution to appear in the amplitudes because, although it may not be obvious, the stated boundary conditions are (particular components of) those in play for the momentum space propagator, from which the amplitude is constructed. We will show this in the following subsection, in which we briefly digress from the master formula in order to investigate how the Volkov wavefunctions arise from worldline path integrals.

\subsection{Mixed boundary conditions and the Volkov wavefunction}
\label{secWavefunction}
Before moving on to the spinor case, we remark that one can, in fact, compute the momentum-space propagator without going \emph{explicitly} via the position-space representation. Returning to the original expression~(\ref{propagator_pos_def}) for $\mathcal{D}_N^{x'x}$, we immediately perform the Fourier transform (\ref{propagator_mom_def}). Now, the exponent $p'\cdot x'-p \cdot x$ in the Fourier kernel is, under the path integral, the same as $p'\cdot x(T)- p\cdot x(0)$, and the spacetime integrals $\ud^4x'\ud^4x$ can be interpreted as $\ud^4 x(T)\ud^4 x(0)$. Hence, taking the Fourier transform of (\ref{propagator_pos_def}) is equivalent to performing a path integral with a free boundary, i.e.~no \emph{apparent} restriction on the endpoints of the worldline. There is though an alternative, but equivalent, perspective; consider the change of the total action, $\delta S$, under the variations of the endpoints of the worldline, $x(0)\rightarrow x(0)+\delta x_0$ and $x(T)\rightarrow x(T)+\delta x_T$:
\begin{align}
    \delta S&\equiv \delta S_{B}+\delta \big( p'.x(T)-p.x(0)\big) \nonumber \\
    &= \left[\frac{1}{2}\dot{x}(0)+a(x(0))-p\right]\cdot\delta x_{0}-\left[\frac{1}{2}\dot{x}(T)+a(x(T))-p'\right]\cdot \delta x_{T}  \label{mixed}
\end{align}
Integrating over $\delta x_{T}$ and $\delta x_{0}$ therefore returns delta functions which impose the vanishing of the terms in square brackets of (\ref{mixed}); these  are \emph{Robin} boundary conditions which relate the worldline end-point momenta $\dot{x}$ to the end-point positions $x$ and the external asymptotic momenta. It follows that the momentum-space propagator can be computed alternatively from the path integral expression
\begin{align}\label{propagator_mom_PI}
    \mathcal{D}^{p'p}_{N} = 
    (-ie)^{N}\!
    \int_{0}^{\infty}\!\ud T \,\e^{-im^{2} T}\,
\int^{\dot{x}(T)2+a(x(T))=2p'}_{\dot{x}(0)+2a(x(0))=2p}\mathcal{D}x(\tau)\, \e^{iS_{\mathrm{B}}[x(\tau)]}\prod_{i=1}^{N} V[\varepsilon_{i}\, k_i]\;.
\end{align}
In the previous section we carried out the Fourier transform of $\mathcal{D}^{x'x}_{N}$ literally, to obtain $\mathcal{D}^{p'p}_{N}$. Expression \eqref{propagator_mom_PI} shows a more `direct' approach to deriving the master formula in \eqref{Master:sQED:momentum-propagator}, through a modification of the boundary conditions on the path integral. This fits in more naturally with the `worldline philosophy' of incorporating all information into the worldline path integral. Note that evaluation of (74) requires a worldline propagator with different boundary conditions. Indeed, this helps explain a puzzle arising in \cite{103} (Section 3, footnote 3), where a version of the momentum space Master Formula was given that involves a Green function with mixed boundary conditions: by expanding about a suitable reference trajectory, (\ref{propagator_mom_PI}) can be cast into a path integral for the fluctuation variable that must satisfy the mixed boundary conditions $\dot{q}(0) = 0 = q(T)$.

This discussion prompts us to study the propagator $\mathcal{D}_{N}^{xp}$ with mixed boundary conditions which, examining (\ref{propagator_mom_PI}), is given by the integral
\begin{align}\label{propagator_mix_PI}
    \mathcal{D}^{xp}_{N} = 
    (-ie)^{N}\!
    \int_{0}^{\infty}\!\ud T \,\e^{-im^{2} T}\,
\int^{x(T)=x}_{\dot{x}(0)+2a(x(0))=2p}\mathcal{D}x(\tau)\, \e^{iS_{\mathrm{B}}[x(\tau)]}\prod_{i=1}^{N} V[\varepsilon_{i}\, k_i]\;.
\end{align}
To see the significance of the mixed propagator, consider the case $N=0$, that is the tree level two-point function for the scalar field, with mixed boundary conditions. In Feynman diagram language, this is just an external leg, Fourier transformed at one end. Taking the momentum at this end onto the mass-shell, i.e.~performing LSZ reduction, we must recover the scalar Volkov wavefunctions. These are solutions of the Klein-Gordon equation in a plane wave background which reduce to $e^{\pm ip.x}$ in the asymptotic past/future and thus represent incoming and outgoing particles in scattering amplitudes.

To confirm this, we first compute the path integral in (\ref{propagator_mix_PI}) for $N=0$ (we drop the product of vertex operators).  We do not dwell on this step;  the entire integral turns out, unsuprisingly given the nature of the Volkov solutions {and hidden Gaussianity of the worldline path integral}, to be equal to its semi-classical value $\exp [i S_{cl}(T)]$, i.e.~the exponential of the classical action evaluated on the classical path obeying the mixed boundary conditions, which is
\begin{align}\label{scl}
 S_{cl}(T) = (p^2-m^2+i0^{\LCp})T-p \cdot x-\int_{x^{\LCp}-4p_-T}^{x^{\LCp}}\! \ud s\, \frac{2p \cdot a(s)-a^2(s)}{4p_\LCm} \;.
\end{align}
The final step is to take $p^2\to m^2$ and identify the on-shell residue via
\be
    \lim_{p^2\to m^2} -i (p^2- m^2+ i 0^\LCp)  \int_{0}^{\infty}\!\ud T \,\e^{-im^{2} T}\, \e^{i S_{cl}(T)} \;.
\ee
Of course it is clear from the preceding calculations how to proceed; we perform the same manipulations as for the master formula, {in particular taking the $T\to\infty$ limit}, immediately finding
\begin{align}
    \lim_{p^2\rightarrow m^2} -i(p^2-m^2)\mathcal{D}^{xp}=  \exp \bigg[-ip\cdot x-i\int_{-\infty}^{x^\LCp}\!\ud s\, \frac{2p \cdot a(s)-a(s)^2}{4p_{\LCm}}\bigg] {\equiv \varphi^\text{\rm in}_p(x)}. \label{VWFfinal}
\end{align}
The right-hand side is precisely the incoming scalar Volkov wavefunction $\varphi^\text{\rm in}_p(x)$ which reduces to $e^{-ip\cdot x}$ in the asymptotic past. A similar amputation of the propagator $\mathcal{D}^{px}_0$ (where the boundary conditions are swapped) yields the outgoing Volkov wavefunctions, i.e.~those which reduce to $e^{+i{\tilde p}^\prime\cdot x}$ in the asymptotic future.  Of course the same procedure can be applied to the spinor propagator, wherein the path integral with mixed boundary conditions produces the spinor Volkov wavefunctions. Worldline path integrals analogous to \eqref{propagator_mix_PI}, with mixed boundary conditions, have also been used before, in a similar context, to recover the exact
solutions of the Klein-Gordon equation in a constant external electromagnetic field~\cite{Rajeev:2021zae}. For numerical studies of open line instantons see~\cite{DegliEsposti:2021its}.

\subsection{Spinor QED}
{Turning to LSZ reduction in spinor QED, we proceed from \eqref{QED:amplitude:start}, writing $\mathcal{S}_N^{x'x}$ in terms of the kernels appearing in (\ref{Master:QED:correlator}) and 
evaluating the $\slashed{\partial}_{x'}$, $\slashed{\partial}_{x}$ derivatives (using integration by parts) in (\ref{QED:amplitude:start}) to find}
\begin{align}
    \mathcal{M}_{Ns's}^{p'p} & = i
    \lim_{p'^2,p^2\rightarrow m^2}
    \int \!
    \ud^{4}x' \ud^{4}x
    \,
    \e^{i\tilde{p}'\cdot x'-ip\cdot x}\bar{u}_{s'}(p')(\slashed{p}'-m)\Bigl\{
    (-\slashed{p}'
    {+\delta \slashed{a}(x^{\prime \LCp})}
    -m)\,\mathcal{K}_N^{x'x}+e\sum_{i=1}^{N}\slashed{\varepsilon}_{i}e^{ik_{i}\cdot x'}\,\mathcal{K}_{N-1}^{x'x}\Bigr\}(\slashed{p}-m)u_{s}(p)\,.
\end{align}
Next, following~\cite{Ahmadiniaz:2021gsd} we use the \emph{on-shell} relation $\bar{u}_{s'}(p')(\slashed{p}'+m)^{-1}=\bar{u}_{s'}(p')(2m)^{-1}$,
(which is allowed since it does not remove the associated pole, or affect the final expression), and likewise for $(\slashed{p}+m)^{-1}u_{s}(p)$ to find
\begin{align}\label{M_N}
    \mathcal{M}_{Ns's}^{p'p} =  i 
    \lim_{p'^2,p^2\rightarrow m^2}
    \frac{1}{2m}\int d^{4}x'd^{4}x\, e^{i\tilde{p}'\cdot x'-ip\cdot x}\,\bar{u}_{s'}(p')
    &(p^{\prime 2}-m^2)\Bigl\{ \Bigl[-1
    {+\frac{1}{2m} \delta \slashed{a}(x^{\prime +}))}
    \Bigr]\,\mathcal{K}_N^{x'x}\notag\\
    & \hspace{4em}+\frac{e}{2m}\sum_{i=1}^{N}\slashed{\varepsilon}_{i}e^{ik_{i}\cdot x'}\,\mathcal{K}_{N-1}^{x'x}\Bigr\}(p^2-m^2)u_{s}(p)\,.
\end{align}
Due to the worldline approach being based on the second order formalism of QED, the exponent under the proper time integral of the spinor amplitude contains the same terms as for the scalar amplitude -- in particular the parameter and proper time integrals produce (free) \textit{scalar propagators}. Hence it suffices to revise the scalar case for this argument. The difference lies in the spin factor of the kernel, the subleading contibutions (those proportional to $\mathcal{K}_{N-1}$), and the $\delta a(x^{\LCp \prime})$ factor from the covariant derivative. However the differences do not impede processing the $T$, and later $\tau_0$, proper time integrals as for scalars. The result is that the LSZ amputation is realised in precisely the same way, by taking  $T,\tau_0\rightarrow \infty$ as in equations (\ref{eqFT})--(\ref{eqAmpp}).  Moreover, after taking the Fourier transform, the conservation of momenta enforced by
$\delta(x^{+}-x'^{+}+2g^{+}+2p'^{+}T)$
sends
\be
    a(x^{\prime +}) \to 
    a\bigl(2Tp^{\prime +}+x^{\LCp}+
    2g^\LCp
    \bigr)\,.
\ee
The LSZ truncation projects onto asymptotic late time, taking $a(x^{\prime +})\to a^\infty$ when $T\rightarrow \infty$, cancelling the field dependent term in square brackets of (\ref{M_N}).  One may then express~\eqref{M_N} in terms of the momentum space kernel
\be
    \mathcal{M}_{Ns's}^{p'p} =  i
    \lim_{p'^2,p^2\rightarrow m^2}\frac{1}{2m} \bar{u}_{s'}(p')(p^{\prime 2}-m^2)\Bigl\{ -\mathcal{K}_N^{{\tilde{p}}'p} +\frac{e}{2m}\sum_{i=1}^{N}\slashed{\varepsilon}_{i}\,\mathcal{K}_{N-1}^{({\tilde{p}}'+k_i)p}\Bigr\}(p^2-m^2)u_{s}(p)\,.
\ee

Now we address the subleading terms. These are seen to have poles not in the required mass-shell $p^{\prime 2} -m^2$, but  rather in $((p'+k_i)^2-m^2)$. Contributions involving these shifted poles hence vanish after taking the on-shell limit of $(p^{\prime 2}-m^2)/((p'+k_i)^2-m^2)$. 
{This is a remarkable generalisation of the vacuum case~\cite{Ahmadiniaz:2021gsd}.} We can be more precise with how this cancellation comes about. In the kernel of the subleading terms, $\mathcal{K}_{N-1}^{(\tilde{p}^{\prime} + k_{i}) p}$, one must first remove an $\varepsilon_i$ and $k_i$, and then replace $a^\infty$ with $a^\infty + k_i$ in ~\eqref{Master:sQED:amplitude}. This operation leaves $\tilde{p}'+K$ invariant, but it does affect the term $\int^\infty_0d\tau \,p'\cdot \delta a(\tau)$, which was convergent as $\tau\rightarrow \infty$, but now produces a rapidly oscillating phase; noting that the proper time integral calculates the Laplace transform of the function $F(T)$ in (\ref{eqFT}), the Abelian final value theorem can be invoked to confirm that the subleading contributions must vanish.

Since the manipulations are  similar to the scalar case, let us simply record the spinor amplitude in its final form as
 \begin{align}\label{Master:QED:amplitude}
 	\mathcal{M}_{Ns's}^{p'p}&= { \sum_{S = 1}^{N} \sum_{\{i_{1} : i_{S}\}} \mathcal{M}_{NS s's}^{\{i_{1}: i_{S}\}p'p} } \\
 {\mathcal{M}_{NS s's}^{\{i_{1}:i_{S}\}p'p} } &= (-ie)^N(2\pi)^3\delta_{\perp,-}({\tilde p}'+K-p)\int_{-\infty}^{\infty}\!\ud x^{\LCp}e^{i(K+p'-p)_\LCp x^{\LCp}}
 \int_{-\infty}^{\infty}\prod_{i=1}^{N} \ud\tau_i\, 
 \delta\bigg(\sum_{j=1}^N\frac{\tau_j}{N}\bigg)
 \\ \nonumber
 &\times 
 \e^{-i\int_{-\infty}^{0}\bigl[2\tilde{p}' \cdot a(\tau)-a^{2}(\tau)\bigr]d\tau-i\int_{0}^{\infty}\bigl[2p'\cdot\delta a(\tau)-\delta a^{2}(\tau)\bigr]\ud\tau-2i\sum_{i=1}^{N}\bigl[\int_{-\infty}^{\tau_i}k_{i} \cdot a(\tau)\ud\tau-i\varepsilon_i\cdot a(\tau_i)\bigr]}\nonumber \\
 &\times e^{i({\tilde p}' +p)\cdot g-i\sum_{i,j=1}^{N}\bigl(\frac{|\tau_i-\tau_j|}{2}k_i \cdot k_j-i\,\mathrm{sgn}(\tau_i-\tau_j)\varepsilon_i \cdot k_j+\delta(\tau_i-\tau_j)\varepsilon_i \cdot \varepsilon_j\bigr)}
 \bigg|_{\varepsilon_{i_{S+1}}...\varepsilon_{i_{N}}}^{\varepsilon_{i_{1}}...\varepsilon_{i_{S}}=0}\nonumber \\
 &\times \frac{1}{2m}\bar{u}_{s'}(p')\mathrm{Spin}(\tilde{f}_{i_{1}:i_{S}})u_{s}(p)\,.\nonumber
 \end{align} 
After LSZ reduction, the argument of the exponential in the spin factor,~\eqref{generating_functional}, takes the following form
\be \label{Grassman_LSZ}
    -\int_{-\infty}^{\infty}\!\ud\tau \, [\eta\cdot f\cdot \eta + {\j}\cdot \eta]-\int_{-\infty}^{\infty}\!\ud\tau\int_{-\infty}^{\infty}\ud\tau'\bigl[\eta\cdot f(\tau)\cdot \mathfrak{G}(\tau,\tau')\cdot \j(\tau')+\tfrac{1}{4}\j(\tau)\cdot\mathfrak{G}(\tau,\tau')\cdot\j(\tau')\bigr]\,;
\ee
the worldline average in the fermion Green function is also now understood to be $T\llangle f\rrangle=\int^\infty_{-\infty}d\tau f(\tau)$. Also, the background gauge potential, $a$, and field strength, $f$, are understood to be functions of
the classical solution $x_{cl}^{\LCp}(\tau)$ as shown in~\eqref{NeumannBC}. Finally, the sums in the first line of (\ref{Master:QED:amplitude}) are -- as usual -- over the assignation of $S$ photons out of $N$ to the spin part of the vertex operator.

\section{Examples}\label{sect:examples}
%
In this section we provide checks on our amplitude Master Formulae (\ref{Master:sQED:amplitude}) and (\ref{Master:QED:amplitude}), showing by comparison with the existing literature that they are consistent with results expected from Furry picture perturbation theory.

\subsection{$N=1$, nonlinear Compton scattering in scalar QED}
The case $N=1$ describes single photon emission from a (scalar) electron in a plane wave background, which is the well-studied process of `nonlinear Compton scattering'. In this case, several parts of the Master Formulae (\ref{Master:sQED:amplitude}) simplify immediately. First, the delta function fixes $\tau_1 = 0$. Next, the gauge field is evaluated as
\begin{align}\label{cases1}
	a(\tau)=
 \begin{cases}
	a(x^{\LCp}+2p^{\LCp}\tau) \;, & \tau<0\\
	a(x^{\LCp}+2p'^{\LCp}\tau)\; , & \tau>0 \;.
	\end{cases}
\end{align}
This form facilitates an easy conversion of integrals over proper time $\tau$ to integrals over lightfront time $x^{\LCp}$, which are expected in the standard formalism (see also \cite{AntonPlane}). Specifically, we can conveniently treat the positive and negative $\tau$ regions separately. The field-dependent terms in the exponent of the master formula then reduce to
\begin{align}
  &-i\int_{-\infty}^{0}\!\ud \tau [2\tilde{p}'\cdot a(\tau)-a^{2}(\tau)]-i\int_{0}^{\infty}\!\ud \tau [2p'\cdot\delta a(\tau)-\delta a^{2}(\tau)]-2i\int_{-\infty}^{0}\!\ \ud\tau\, k_{1}\cdot a(\tau)\\
&=-i\int_{-\infty}^{x^{\LCp}}\!\ud s^\LCp\, \frac{2p\cdot a(s^{\LCp})-a^2(s^{\LCp})}{2p^{\LCp}} - i\int_{x^{\LCp}}^{\infty}\!\ud s^{\LCp}\, \frac{2p'\cdot\delta a(s^{\LCp})-\delta a^2(s^{\LCp})}{2p'^{\LCp}} \;,
\end{align}
in which we simply inserted (\ref{cases1}) and used momentum conservation in the transverse directions to eliminate $k_1$ in favour of $p'$ and $p$. {With this, expanding (\ref{Master:sQED:amplitude}) for $N=1$ to linear order in $\varepsilon_1$, and using the Fourier representation of the momentum conserving $\delta$-functions shows that the amplitude is equivalent to} 
\begin{align}
    \label{N1scalar-checked}
	\mathcal{A}^{p'p}_{1}=-ie\int \ud^{4}x\, \left\{\tilde{p}'_{\mu}+p_{\mu}-2a_{\mu}(x^{\LCp})\right\}\varepsilon^{\mu}_1e^{ik_1.x}\varphi_{p'}^{\text{out}}(x)\varphi^{\text{\rm in}}_{p}(x)\,,
\end{align}
where $\varphi^{\text{\rm in}}_{p}$ is the incoming scalar Volkov wavefuncion of (\ref{VWFfinal}) while $\varphi^{\text{\rm out}}_{p'}$ is the outgoing wavefunction,
\begin{align}\label{sVolkov}
	\varphi^{\text{out}}_{p'}(x)=
    {\e^{i\tilde{p}' \cdot x}}\exp\left[-i\int_{x^{\LCp}}^{\infty}\!\ud s^{\LCp}\, \frac{2p' \cdot \delta a(s^{\LCp})-\delta a^2(s^{\LCp})}{2p'^{\LCp}}\right].
\end{align}
Expression (\ref{N1scalar-checked}) is precisely the expected result for nonlinear Compton scattering in scalar QED, providing a positive check on our master formula.

We stress that the method we employed above to process the worldline integrals was meant only to allow direct comparison with existing results. It is \emph{not} the approach we wish to take in future work; instead, we will use the worldline representation to deal \emph{directly} with the $\tau$-integrals. {Since the major advantages of the worldline approach include that (a) one does not have to split amplitudes into sectors according to permutations of external legs, and (b) internal momentum integrals are recast in terms of the proper time integral,} we expect this to provide some advantage over the standard formalism, at least in various physical limits of interest. This will be discussed elsewhere.

\subsection{$N=1$, nonlinear Compton scattering in spinor QED}
Let us now confirm the $N=1$ case for spinor QED, {which requires expanding the Master Formula (\ref{Master:QED:amplitude})} to linear order in $\varepsilon_{1}$. Since the field dependence of the exponent in for spinor QED contains that of scalar QED one may write the resulting amplitude using the scalar Volkov wavefunctions,~\eqref{sVolkov}, as
\be\label{amp_spin_N1}
    \mathcal{M}_{1s's}^{p'p}=-ie\frac{1}{2m}\int d^{4}x\, e^{ik_1\cdot x}\varphi_{p'}^{\text{out}}(x)\varphi_{p}^{\text{\rm in}}(x)\bar{u}_{s'}(p')\Bigl[(\tilde{p}'+p-2a(x^{\LCp}))\cdot\varepsilon_1\mathrm{Spin}(\emptyset)+\mathrm{Spin}(\tilde{f_1})\Bigr]u_{s}(p)\,,
\ee
requiring only the evaluation of the spin factor (we have again used the Fourier representation of the $\delta$-functions). Before embarking upon the comparison to the standard formalism, we should emphasise that the approach outlined here, namely writing in terms of spacetime averages with steps to follow, is necessary to make the connection to the perturbative Furry picture with Volkov wavefunctions. However, this would be inefficient for practical worldline calculations. 

{The spin factors are determined using (\ref{W_0}) and (\ref{W_f_1}) under the LSZ reduction (\ref{Grassman_LSZ})} and the inverse symbol map,~\eqref{symbolic_map}. {Because of the nilpotency of $f$ one has, \emph{under the inverse symbol map},   $\exp(-\int^\infty_{-\infty}\!\ud\tau \, \eta\cdot f\cdot \eta) = 1-\int^\infty_{-\infty}\!\ud\tau \eta \cdot f \cdot \eta$, and therefore the factor without photon insertion is readily determined to be}
\be
    \mathrm{Spin}(\emptyset)=\Bigl[1-\frac{1}{2p^{\prime +}}\slashed{n}\delta\slashed{a}(x^{\LCp})\Bigr]\Bigl[1+\frac{1}{2p^{+}}\slashed{n}\slashed{a}(x^{+})\Bigr]\,,
\ee
where we have already transformed the parameter integral to a spacetime average and computed its value. This is simply the Dirac matrix structure necessary to construct the spinor Volkov wavefunctions. 

Let us next treat the single photon spin factor, $\mathrm{Spin}(\tilde{f}_{1})$. Beginning with the Grassmann integral with one photon insertion, provided in~\eqref{W_f_1} we apply the inverse symbolic map in~\eqref{symbolic_map} and realise the LSZ reduction according to~\eqref{Grassman_LSZ}. The various worldline averages are then transformed into their corresponding spacetime averages as was done in the $N=1$ scalar case, to find
\begin{align}\label{spin_f}    \mathrm{Spin}(\tilde{f}_{1})&= -\frac{1}{2}[\slashed{k}_{1},\slashed{\varepsilon}_{1}] + k_{1}^{+}\varepsilon_{1}\cdot\Bigl(-\frac{\delta a(x^{+})}{2p^{\prime+}}+\frac{a(x^{+})}{2p^{+}}\Bigr)+\varepsilon_{1}\cdot\Bigl(\frac{\delta a(x^{+})}{2p^{\prime+}}+\frac{a(x^{+})}{2p^{+}}\Bigr)\frac{1}{2}[\slashed{k}_{1},\slashed{n}]\nonumber \\
    &+k_{1}^{+}\frac{1}{2}\Bigl[\slashed{\varepsilon}_{1},\frac{\delta\slashed{a}(x^{+})}{2p^{\prime+}}+\frac{\slashed{a}(x^{+})}{2p^{+}}\Bigr]+\Bigl[k_{1}\cdot\Bigl(\frac{\delta a(x^{+})}{2p^{\prime+}}+\frac{a(x^{+})}{2p^{+}}\Bigr)+2k_{1}^{+}\Bigl(\frac{\delta a(x^{+})}{2p^{\prime+}}\cdot\frac{a(x^{+})}{2p^{+}}\Bigr)\Bigr]\slashed{n}\slashed{\varepsilon}_{1}\nonumber\\
    &+\frac{2k_{1}^{+}}{2p^{\prime+}2p^{+}}\varepsilon_{1}\cdot\big[a(x^{+})\delta\slashed{a}(x^{+})+\delta a(x^{+})\slashed{a}(x^{+})\big]\slashed{n}+(k_{1}+a^\infty)_{\mu}\varepsilon_{1\nu}n_{\alpha}\Bigl(\frac{\delta a(x^{+})}{2p^{\prime+}}-\frac{a(x^{+})}{2p^{+}}\Bigr)_{\beta}i\gamma_{5}\epsilon^{\mu\nu\alpha\beta}\,.
\end{align}
Next, we express the photon momentum, $k_1$, in terms {of the electron momenta and asymptotic value of the background field. For the  ${+, \perp}$ components we can use momentum conservation, $k_{1}^{+, \perp} = (p -\tilde{p}')^{+,\perp}$. } {The $k_{1}^{-}$ component requires us to carry out an an integration by parts with respect to $x^{+}$. We illustrate this step, to be applied to the various $\slashed{k}_1$ terms in (\ref{spin_f}), with the following manipulation:
\begin{equation}
    \int\!\ud^{4}x\,
    \e^{ik_1\cdot x}{k_{1}^\mu}\varphi_{p'}^{\text{out}}(x)\varphi_{p}^{\text{\rm in}}(x)
    =
    \int\!\ud^{4}x\,
    \e^{ik_{1}\cdot x}\Bigl[\Bigl(\frac{2p\cdot a(x^{+})-a(x^{+})^{2}}{2p^{+}}-\frac{2p'\cdot\delta a(x^{+})-\delta a(x^{+})^{2}}{2p^{\prime+}}\Bigr){n^\mu}+{p^\mu}-{\tilde{p}^{\prime\mu}}\Bigr]\varphi_{p'}^{\text{out}}(x)\varphi_{p}^{\text{\rm in}}(x)\,,
    \label{eqIBP}
\end{equation}
In fact, if additional factors of $a(x^{+})$ appear under the above integral, in turns out that the additional derivatives produced by integrating by parts \textit{always} contract away. Therefore (\ref{eqIBP}) can be used throughout (\ref{spin_f}). Moreover, applying the above procedure to $k_1$ in the $\gamma_5$ term of~\eqref{spin_f}, one can see that in effect $k_1^{\mu}\to p^{\mu}-\tilde{p}^{\prime \mu}$, since the two $n^{\mu}$ contract to zero against the Levi-Civita tensor. In fact the only term in which the $n^{\mu}$ part of (\ref{eqIBP}) survives after these replacements is the first term on the RHS of (\ref{spin_f}).

Last, since we are taking the on-shell limit we may use the Dirac equation for the sandwiching spinors so as to send their corresponding $\slashed{p}$ and $\slashed{p}'$ to $m$, anti-commutating where necessary. Again, illustrating this step with the $\gamma_{5}$ term in (\ref{spin_f}) we rewrite $\gamma_5$ in terms of products of four matrices using~\eqref{symbolic_map}. After acting on the spinor solutions at most three matrices will remain. After this process, the $\gamma_5$ term, as it appears in the amplitude~\eqref{amp_spin_N1}, becomes
\begin{align}
    (k_{1}+a^\infty)_{\mu}\varepsilon_{1\nu}n_{\alpha}\Bigl(\frac{\delta a(x^{+})}{2p^{\prime+}}-&\frac{a(x^{+})}{2p^{+}}\Bigr)_{\beta}i\gamma_{5}\epsilon^{\mu\nu\alpha\beta} =
    (p^{+}+p'^{+})\frac{1}{2}\Bigl[\frac{\delta\slashed{a}(x^{+})}{2p^{\prime+}}-\frac{\slashed{a}(x^{+})}{2p^{+}},\slashed{\varepsilon}_1\Bigr]+(p+p')\cdot\varepsilon_1\slashed{n}\Bigl(\frac{\delta\slashed{a}(x^{+})}{2p^{\prime+}}-\frac{\slashed{a}(x^{+})}{2p^{+}}\Bigr)\nonumber\\
    &+(p+p')\cdot\Bigl(\frac{\delta a(x^{+})}{2p^{\prime+}}-\frac{a(x^{+})}{2p^{+}}\Bigr)\slashed{\varepsilon}_1\slashed{n}-m\Big\{\slashed{\varepsilon}_1,\slashed{n}\Bigl(\frac{\delta\slashed{a}(x^{+})}{2p^{\prime+}}-\frac{\slashed{a}(x^{+})}{2p^{+}}\Bigr)\Big\}\,.
\end{align}
Using the above steps to replace $k_{1}^{\mu}$ in the remaining terms of (\ref{spin_f}), after some algebra one may gather terms to find that
\be
   \bar{u}_{s'}(p^{\prime}) \Bigl\{ (\tilde{p}'+p-2a(x^{+}))\cdot\varepsilon_1\mathrm{Spin}(\emptyset)+\mathrm{Spin}(\tilde{f_1}) \Bigr\} u_{s}(p) = 2m\bar{u}_{s'}(p^{\prime}) \Bigl\{\slashed{\varepsilon}_1-\frac{1}{2p^{\prime+}}\slashed{n}\delta\slashed{a}(x^{+})\slashed{\varepsilon}_{1}+\frac{1}{2p^{+}}\slashed{\varepsilon}_{1}\slashed{n}\slashed{a}(x^{+})\Bigr\}u_{s}(p)\,,
\ee
and hence
\be
    {\mathcal{M}^{p'p}_{1 s' s}}	=-ie \int\!\ud^{4}x\, \e^{ik\cdot x}\Psi_{p',s'}^{\text{out}}(x)\slashed{\varepsilon}_1\Psi_{p,s}^{\text{\rm in}}(x)\,,
\ee
where we have used the spinor Volkov wavefunctions, which read
\begin{align}
    \Psi_{p,s}^{\text{in}}(x)	&=\Bigl[1+\frac{1}{2p^{+}}\slashed{n}\slashed{a}(x^{+})\Bigr]u_{s}(p)\varphi_{p}^{\text{\rm in}}(x)\,,\\
    \Psi_{p',s'}^{\text{out}}(x)	&=\bar{u}_{s'}(p')\Bigl[1-\frac{1}{2p^{\prime +}}\slashed{n}\delta\slashed{a}(x^{+})\Bigr]\varphi_{p'}^{\text{out}}(x)\,.
\end{align}
This successfully verifies that the worldline approach reproduces the known amplitude for the $N = 1$ process.

\subsection{$N=2$, double nonlinear Compton scattering in scalar QED}
To complete our discussion of the relevant structures in scalar QED we must also consider the case $N=2$, where the so-called seagull vertex (the four-point scalar-photon-photon-scalar vertex) first appears.
We will describe the way this works briefly here, as the calculations proceed largely as for $N=1$, leaving the details for the Appendix. Expanding (\ref{Master:sQED:amplitude}), there are now two $\tau$-integrals, with one, say $\tau_2$, fixed by the worldline delta function in \eqref{Master:sQED:amplitude}, and the other, say $\tau_1$, remaining. The mapping onto Feynman diagrams is most natural: the contributions from $\tau_1>0$ and $\tau_1<0$ recover one each of the expected contributions from the two diagrams with two 3-point vertices, with $\tau_1$ being mapped to the lightfront time of one vertex. The seagull contribution is picked up from the term in \eqref{Master:sQED:amplitude} which goes like $\varepsilon_1\cdot \varepsilon_2$; this comes with a delta function with support at exactly $\tau_1=0$, hence leaving only a single unevaluated integral, as expected. The full calculation is presented in Appendix~\ref{app:N2}.

\section{Conclusions}\label{sect:conclusions}
We have presented worldline Master Formulae for all-multiplicity tree level scattering amplitudes of 2 massive charged particles and $N$ photons, in a plane wave background, in both scalar and spinor QED. The background field may have arbitrary strength and functional profile, and is treated without approximation throughout. This is particularly relevant as the target application of our results is to laser-matter interactions in the \emph{high intensity} regime~where the field is characterised by a dimensionless strength (the coupling to matter) larger than unity, and hence must be treated without recourse to perturbation theory.

Our Master Formulae have been derived using the worldline approach to quantum field theory. While several previous publications have derived wordline Master Formulae for various correlation functions in vacuum, or even at higher loop level in background fields, our focus here has been on scattering amplitudes involving external matter. As such it was necessary to identify the \emph{worldline description} of LSZ reduction in a plane  wave background. We found this to be a fairly direct generalisation of the known worldline prescription for LSZ  amplitudes in vacuum~\cite{Bonocore:2020xuj,Mogull:2020sak}. {A second notable generalisation from known results in vacuum holds for the spinor case: namely that in the second order formalism, which implies a split into `leading' and `subleading' terms, {only the former} survives the on-shell limit {once the LSZ prescription is imposed}. Furthermore, the background field dependent part of this leading term \textit{also} drops out in the asymptotic limit. This allows for a large number of terms to be discarded (and in the vacuum case allowed for the gauge invariance of the amplitudes to be manifest).}

We have checked our results against the existing literature, which contains only \emph{low}-multiplicity amplitudes derived using Feynman rules. Explicitly, these are the cases $N=1$ and $N=2$, or single and double nonlinear Compton scattering. Moving beyond scattering amplitudes, we have also seen how to recover \emph{off-shell} quantities, in particular the scalar and spinor {correlation functions dressed by the background
and the Volkov wavefunctions}, from worldline path integrals. The latter is a particularly interesting case as it exposes the relevance of mixed boundary conditions; the relevant path integrals carry Dirichlet conditions at one limit, representing the local spacetime argument of the wavefunction, and Robin boundary conditions at the other limit, encoding the asymptotic momentum characterising the Volkov solution. 

It is fair to say that the Master Formulae for amplitudes we have derived here still require, for a chosen number of photons $N$, some processing in order to extract {all their} physical content. In future work we will pursue methods of evaluating the remaining {proper time} integrals in an efficient manner, or in an approximate manner relevant to interesting physical regimes. {Here, benefit should be gained by \textit{not} breaking the parameter integrals into ordered sectors corresponding to photon permutations, which will maximally exploit the calculational efficiency}.  Constructing observables from our amplitudes at $N>2$ (which are lacking in the literature) will help to benchmark numerical codes which approximate multi-photon processes using sequential single-photon emissions. It would be revealing to compare our expressions with the compact all-multiplicity results of~\cite{Adamo:2020syc,Adamo:2020yzi}. We also plan to generalise our results to higher loop orders, in order to pursue the Ritus-Narozhny conjecture on the behaviour of loop corrections at very high intensity, see~\cite{Fedotov:2016afw,Fedotov:2022ely} for reviews.

\begin{acknowledgments}
The authors are supported by the EPSRC Standard Grants  EP/X02413X/1 (PC, JPE) and EP/X024199/1 (AI, KR), and the STFC consolidator grant ST/X000494/1 (AI).

\end{acknowledgments}
\appendix

\section{Master formula check for $N=2$}\label{app:N2}
In this appendix we confirm that the master formula (\ref{Master:sQED:amplitude}) correctly reproduces, at $N=2$, the amplitude for `double nonlinear Compton scattering'~\cite{Seipt:2012tn,Mackenroth:2012rb} in scalar QED, that is the emission of two photons from a particle in a plane wave background. (By crossing symmetry this is directly related to the amplitude for the Compton effect in the background.) Recall that in scalar QED, the standard approach would require evaluation of three separate Feynman diagrams -- conveniently combined into one calculation on the worldline -- one of which contains the four-point seagull vertex.

Starting from (\ref{Master:sQED:amplitude}) with $N=2$, the LSZ factor $\delta(\tau_1/2+\tau_2/2)$ means that we have only one non-trivial proper-time integral, over, say, $\tau_1$. 
It is convenient to split this integral into three pieces and analyse each separately; we split the integration range into $-\infty < \tau_1 < 0^\LCm$, $0^\LCm < \tau_1 < {0^\LCp}$ and  $0^\LCp < \tau_1 < \infty$, and refer henceforth to the corresponding contribution to the amplitudes as $\mathcal{A}^{p'p}_{2-}$,  $\mathcal{A}^{p'p}_{2\delta}$ and  $\mathcal{A}^{p'p}_{2+}$, respectively. 

\subsection{$\tau_1\in(0,\infty)$}
When $\tau_1>0$, the field-independent terms in the exponential of (\ref{Master:sQED:amplitude}) reduce to
{\begin{align}\label{no_a_terms_N=2}
	i(\tilde{p}'+p)\cdot (k_1-k_2)\tau_1+\varepsilon_{1}\cdot(\tilde{p}'+p-k_2)+\varepsilon_{2}\cdot(\tilde{p}'+p+k_1)-2i\tau_1k_1\cdot k_2+i\left(K_{\LCp}+p'_{\LCp}-p_{\LCp}\right)x^{\LCp} \,.
\end{align} }
The gauge field at the interaction points $\pm\tau_1$ ({indicating the insertion point of photon with momentum $k_{1}$}) takes the values
{\begin{align}
a(\tau_1)&= a(x^{\LCp}+ \tau_1(2p'^\LCp + k_1^\LCp-k_2^\LCp)\,,\\
{a(-\tau_1)}&=a(x^{\LCp}- \tau_1(2p^\LCp + k_1^\LCp-k_2^\LCp)\,.
\end{align}}
This motivates us to make the change of variable {$x^{\LCp} \to x^{\LCp}-\tau_1(2p^\LCp+k_1^\LCp-k_2^\LCp)$},
such that the field-independent terms (\ref{no_a_terms_N=2}) transform to 
\begin{align}\label{free_terms_simplified}
\mathcal{T}_0&\equiv i\left(4(p_++k_{1+})q^{\LCp}-2q^2_{\perp}-2m^2+i0^{\LCp}\right)\tau_1+\varepsilon_1\cdot(2\tilde{p}'+k_1)+\varepsilon_2\cdot(\tilde{p}'+p+k_1)+i\left(K_{\LCp}+p'_{\LCp}-p_{\LCp}\right)x^{\LCp}\\\nonumber
&-i(2p'+a^\infty)a^\infty\tau_1    \,,
\end{align}
where we have defined $q=p-k_2$ and used the fact the momenta are on-shell to simplify. We shall shortly need the last term $-i(2p'+a^\infty)a^\infty\tau_1$ to simplify some of the field dependent terms. Before going into that, we return to the exponent of (\ref{Master:sQED:amplitude}) and note that the following field-dependent term is already sufficiently simplified.
\begin{align}
\mathcal{T}_{1}&\equiv  -2 \sum_{i=1}^N\varepsilon_i\cdot a(\tau_i)\rightarrow-2\varepsilon_1\cdot a(x^{\LCp})-2\varepsilon_2 \cdot a(x^{\LCp}+4q^{\LCp}\tau_1)\,.
\end{align}
The rest of the field-dependent terms combine with $-i(2p'+a^\infty)a^\infty\tau_1$ from \eqref{free_terms_simplified} to yield
\begin{align}
    \mathcal{T}_2-&i(2p'+a^\infty)a^\infty\tau_1\equiv\\
    \nonumber &-2i\sum_{i=1}^N\int_{-\infty}^{\tau_i}\! \ud \tau k_i\cdot a(\tau)
    -i\int_{-\infty}^{0}\!\ud\tau [2\tilde{p}'\cdot a(\tau)-a^{2}(\tau)]
    -i\int_{0}^{\infty}\!\ud\tau [2p'\cdot \delta a(\tau)-\delta a^{2}(\tau)]
    -i(2p'+a^\infty)\cdot a^\infty\tau_1\\\nonumber
    &=-2i\sum_{i=1}^N\int_{-\infty}^{\tau_i}\!\ud\tau \, k_i\cdot a(\tau)
    -i\int_{-\infty}^{\tau_1}\!\ud\tau\, [2\tilde{p}'\cdot a(\tau)-a^{2}(\tau)]
    -i\int_{\tau_1}^{\infty} \!\ud\tau [2p'\cdot \delta a(\tau)-\delta a^{2}(\tau)] \;.
\end{align}
We now use the {dependence of $a_{\mu}(x_{\rm{cl}}(\tau)$ on the} classical solution to transform the proper-time integrals into space-time integrals and simplify the above terms as
\begin{align}
    &-2i\sum_{i=1}^N\int_{-\infty}^{\tau_i}\! \ud\tau \, k_i\cdot a(\tau)
    -i\int_{-\infty}^{\tau_1}\!\ud\tau\, [2\tilde{p}'\cdot a(\tau)-a^{2}(\tau)]-i\int_{\tau_1}^{\infty} \!\ud\tau\, [2p'\cdot \delta a(\tau)-\delta a^{2}(\tau)]\\
    &=-i\int_{-\infty}^{x^{\LCp}}\frac{2 p.a(s)-a^2(s)}{2p^\LCp}ds-i\int_{x^{\LCp}}^{x^{\LCp}+4q^{\LCp}\tau_1} \!\ud s\, \frac{2 q\cdot a(s)-a^2(s)}{2q^\LCp}-i\int_{x^{\LCp}+4q^{\LCp}\tau_1}^{\infty}\!\ud s\, \frac{2 p'\cdot \delta a(s)-\delta a^2(s)}{2p'^\LCp}\,,
\end{align}
where we have used momentum conservation to replace $\tilde{p}_{\perp}+K_{\perp}$ with $p_{\perp}$, and $\tilde{p}_{\perp}+k_{1\perp}$ with $q_{\perp}$. The contribution $\mathcal{A}^{p'p}_{2+}$ to the amplitude from $\tau_1>0$ can then be written as
\begin{align}\label{amplitude_positive}
    \mathcal{A}^{p'p}_{2+}=2(-ie)^2(2\pi)^3\delta_{\perp,-}\left(\tilde{p}'+K-p\right)\int_{-\infty}^{\infty}\!\ud x^{\LCp}\int_{0}^{\infty}\ud \tau_1 \, \left. \e^{\mathcal{T}_0+\mathcal{T}_1+\mathcal{T}_2}\right\rvert_{\textrm{lin. }\varepsilon}
\end{align}
We are now going to show that the right hand side of the above expression is equivalent to {one of the three Feynman diagram contributions to double nonlinear Compton, namely that containing two 3-point vertices in which photon $k_1$ is emitted on the outgoing leg. The Feynman rules give this contribution as}
\begin{align}
    &(-ie)^2\int\!\ud^4x'\ud^4 x\,  \e^{i k_1\cdot x'}\Big[\varphi^{\rm out}_{p'}(x')\,(\varepsilon_1\cdot \stackrel{\leftrightarrow}{D_{x'}}) \,G(x',x)(\varepsilon_2 \cdot\stackrel{\leftrightarrow}{D_x})\,\varphi^{\rm in}_{p}(x)\Big]e^{ik_2 \cdot x}\,,
\end{align}
where $D$ denotes the {background-covariant} derivative and $G(x',x)=\mathcal{D}^{x'x}_{0}$ {is the scalar particle propagator in the plane wave background} {(the double arrow indicates the right$-$left alternating derivative)}. {We then observe that this is equivalent to}
\begin{align}
\int\!\ud^4x'\ud^4 x\, \varphi^{\rm out}_{p'}(x'-i\varepsilon_1)\,e^{i k_1\cdot x'-2\varepsilon_1 \cdot a(x')}\,G(x'+i\varepsilon_1,x-i\varepsilon_2)\,e^{ik_2\cdot x-2\varepsilon_2\cdot a(x)}\,\varphi^{\rm in}_{p}(x+i\varepsilon_2) \bigg|_{\textrm{lin. }\varepsilon_{1} \ldots \varepsilon_{N}} \;.
\end{align}
{Taking this expression, we start by using the Fourier representation of $G(x',x)$ to rewrite it as} 
\begin{align}
   &\int\! \ud^4x' \,\ud^4x\, \varphi^{\rm out}_{p'}(x'-i\varepsilon_1)\, 
   \e^{i k_1 \cdot x'-2\varepsilon_1\cdot a(x')}\,G(x'+i\varepsilon_1,x-i\varepsilon_2)\,
   \e^{ik_2\cdot x-2\varepsilon_2\cdot a(x)}\,\varphi^{\rm in}_{p}(x+i\varepsilon_2)  \\\nonumber
   =&\int\!\frac{\ud^4r}{(2\pi)^4}\ud^4x'\, \ud^4x\, \varphi^{\text{out}}_{p'}(x'-i\varepsilon_1)\,e^{i k_1\cdot x'-2\varepsilon_1\cdot a(x')}\frac{i\e^{-ir\cdot (x'-x+i\varepsilon_1+i\varepsilon_2)-i\int_{x^{\LCp}}^{x'^{\LCp}}\frac{2r\cdot a(s)-a^2(s)}{4r_-}\ud s}}{r^2-m^2+i0^{\LCp}}\,e^{ik_2\cdot x-2\varepsilon_2 \cdot a(x)}\,\varphi^{\text{\rm in}}_{p}(x+i\varepsilon_2)
\end{align}
We can easily evaluate the $x'^{\LCm,\LCperp},x^{\LCm,\LCperp}$ and $r^{\LCm,\LCperp}$ integrals and rewrite the propagator denominator using a standard Schwinger proper-time integral to obtain
\begin{align}
    &(2\pi)^3\delta_{\perp,-}(\tilde{p}'+K-p)
    \e^{p\cdot \epsilon_1+ q.\varepsilon_2}\int_{-\infty}^{\infty} \!\ud x'^{\LCp}\,\,
    \e^{i(p_++k_{1+}-r_+)x'^{\LCp} -2\varepsilon_1\cdot  a(x'^{\LCp})}
    \e^{-i\int_{x'^{\LCp}}^{\infty}\frac{2p'\cdot \delta a(s)-\delta a^2(s)}{2p^{\LCp}}ds}\\\nonumber  
    &\times 2\int_{-\infty}^{\infty}\!\ud x^{\LCp} \e^{-2\varepsilon_2\cdot a(x^{\LCp})}e^{-ix^{\LCp}q_{\LCp}}\int_{0}^{\infty}d\tau_1 \int \frac{\ud r_\LCp}{2\pi}\,
    \e^{ir_+(x^{\LCp}-x'^{\LCp}+4q^{\LCp}\tau_1)}e^{-2i\tau_1\left[q_{\perp}^2+m^2-i0^{\LCp}\right]}\\\nonumber
    &\times \e^{-i\int_{x^{\LCp}}^{x'^{\LCp}}\!\ud s\, \frac{2q\cdot a(s)-a^2(s)}{2q^{\LCp}}-i\int_{-\infty}^{x^{\LCp}}\!\ud s\,\frac{2p\cdot a(s)-a^2(s)}{2p^{\LCp}}} \;.
\end{align}
{The $r_\LCp$ integral can now be evaluated to give $2\pi\delta(x^{\LCp}-x'^{\LCp}+8q_{\LCm}\tau_1)$. The remaining $x'^{\LCp}$ integral is therefore trivialised and effects the replacement $x'^{\LCp}\rightarrow x^{\LCp}+8q_{\LCm}\tau_1$. Taking the multi-linear limit, one recovers precisely the right hand side of~\eqref{amplitude_positive} as promised.}  

\subsection{$\tau_1\in(-\infty,0^{-})$}
%
{For $\tau_1<0$, one recovers the Feynman diagram contribution in which photon $k_2$ is emitted from the outgoing leg. The proof of this follows exactly the same steps as for $\mathcal{A}^{p'p}_{2+}$ above. Hence we simply state that}
\begin{align}\label{NLC2-2}
    \mathcal{A}^{p'p}_{2-}=(-ie)^2\int\! \ud^4x' \ud^4x\, 
    \e^{i k_2 \cdot x}\Big[\varphi^{\rm out}_{p'}(x')\,(\varepsilon_2\cdot\stackrel{\leftrightarrow}{D_{x'}}) \,G(x',x)(\varepsilon_1\cdot\stackrel{\leftrightarrow}{D_x})\,\varphi^{\rm in}_{p}(x)\Big] \e^{ik_1 \cdot x} \;.
\end{align}
%
\subsection{$\tau_1\in(0^{-},0^{\LCp})$}
%
In this range, the field-independent term in the exponent of (\ref{Master:sQED:amplitude}) going like $\delta(\tau_1)\epsilon_1 \cdot\epsilon_2$ cannot be neglected. Noting that this term is already linear in both $\epsilon_{1}$ and $\epsilon_{2}$, the corresponding contribution to the amplitude is immediately seen to be proportional to the $\tau_1\rightarrow 0$ and $\epsilon_{1,2}\rightarrow 0$ limit of the integrand of the proper-time integral:  
\begin{align}
 	\mathcal{A}^{p'p}_{2\delta}&=-2(-ie)^2(2\pi)^3\delta_{\LCperp,\LCm}\left(\tilde{p}'+K-p\right)\\\nonumber
	&\times\int_{-\infty}^{\infty}dx^{\LCp}(i\varepsilon_1\cdot\varepsilon_2)e^{+i\left(K+p'-p\right)_{\LCp}x^{\LCp}-i\int_{-\infty}^{0}[2\tilde{p}'\cdot a(\tau)-a^{2}(\tau)]\ud\tau-i\int_{0}^{\infty}[2p'\cdot\delta a(\tau)-\delta a^{2}(\tau)]\ud\tau-2i\int_{-\infty}^{0}K \cdot a(\tau)\ud\tau}
 \end{align}
By inspection, this is equivalent to
 \begin{align}\label{NLC2-3}
	\mathcal{A}^{p'p}_{2\delta} = -2i(-ie)^2\varepsilon_1 \cdot \varepsilon_2\int\!\ud^4x\, \e^{i(k_1+k_2)\cdot x}\varphi^{\rm out}_{p'}(x)\varphi^{\rm in}_{p}(x) \;,
\end{align}
{which is indeed the seagull vertex contribution to double nonlinear Compton scattering. Summing (\ref{amplitude_positive}), (\ref{NLC2-2}) and (\ref{NLC2-3}) recovers the full amplitude.}

\bibliography{LSZbib}
\end{document}